\documentclass[a4paper,12pt]{article}

\pdfoutput=1 % if your are submitting a pdflatex (i.e. if you have
\usepackage{cite}
\usepackage{jheppubmod}
\usepackage{enumitem}
\usepackage{amsmath}
\usepackage{physics}
\usepackage{graphicx}
\usepackage{amsfonts} 
\usepackage{amssymb}
\usepackage[dvipsnames, svgnames, x11names ]{xcolor}
\usepackage{color}
\usepackage{braket}
\graphicspath{ {./1 /} }
\usepackage{subcaption}
\usepackage{graphicx}% Include figure files
\usepackage[T1]{fontenc} % if needed
\usepackage{tikz}
\usepackage{physics}
\usepackage{amsmath}
\usepackage{dsfont}
\newtheorem{theorem}{Theorem}
\usepackage{graphicx}
\usepackage{bbold}
\graphicspath{ {./1 /} }

\usepackage{tikz}
\usepackage{amsmath} % for \text
\usepackage{mathrsfs} % for \mathscr
\usepackage{xfp} % higher precision (16 digits?)
\usepackage[outline]{contour} % glow around text
\usetikzlibrary{decorations.markings,decorations.pathmorphing}
\usetikzlibrary{angles,quotes} % for pic (angle labels)
\usetikzlibrary{arrows.meta} % for arrow size
\contourlength{1.4pt}

\tikzset{>=latex} % for LaTeX arrow head
\colorlet{myred}{red!80!black}
\colorlet{myblue}{blue!80!black}
\colorlet{mygreen}{green!80!black}
\colorlet{mydarkgreen}{green!30!black}
\colorlet{mydarkred}{red!50!black}
\colorlet{myveryred}{red!30!black}
\colorlet{mydarkblue}{blue!50!black}
\colorlet{mylightblue}{mydarkblue!6}
\colorlet{mypurple}{blue!40!red!80!black}
\colorlet{mydarkpurple}{blue!40!red!50!black}
\colorlet{mylightpurple}{mydarkpurple!80!red!6}
\colorlet{myorange}{orange!40!yellow!95!black}
\tikzstyle{cone}=[mydarkblue,line width=0.2,top color=blue!60!black!30,
                  bottom color=blue!60!black!50!red!30,shading angle=60,fill opacity=0.9]
\tikzstyle{cone back}=[mydarkblue,line width=0.1,dash pattern=on 1pt off 1pt]
\tikzstyle{world line}=[myblue!60,line width=0.4]
\tikzstyle{world line t}=[mypurple!60,line width=0.4]
\tikzstyle{particle}=[mygreen,line width=0.5]
\tikzstyle{photon}=[-{Latex[length=4,width=3]},myorange,line width=0.4,decorate,
                    decoration={snake,amplitude=0.9,segment length=4,post length=3.8}]
\tikzstyle{singularity}=[myred,line width=0.6,decorate,
                         decoration={zigzag,amplitude=2,segment length=6.17}]
\tikzset{declare function={%
  penrose(\x,\c)  = {\fpeval{2/pi*atan( (sqrt((1+tan(\x)^2)^2+4*\c*\c*tan(\x)^2)-1-tan(\x)^2) /(2*\c*tan(\x)^2) )}};%
  penroseu(\x,\t) = {\fpeval{atan(\x+\t)/pi+atan(\x-\t)/pi}};%
  penrosev(\x,\t) = {\fpeval{atan(\x+\t)/pi-atan(\x-\t)/pi}};%
  kruskal(\x,\c)  = {\fpeval{asin( \c*sin(2*\x) )*2/pi}};% Penrose coordinates for Kruskal
}}

\def\Nsamples{20} % number samples in plot

\def\R{0.08} % size lightcone
\newtheorem{definition}{Definition}[section]
\newtheorem{lemma}[theorem]{Lemma}

\title{\boldmath Algebraic traversable wormholes }

\author{Eyoab Bahiru}
\affiliation{Department of Physics, Technion,\\
Haifa 32000, Israel}

\emailAdd{joabdej@gmail.com}

\abstract{We discuss a large $N$ limit for the thermofield double state which at the extreme ($N=\infty$) limit is dual in the bulk to a back-reacted traversable wormhole, by making use of an operator in the algebra at infinity, an algebra familiar in the literature from the study of quasi-local algebras. We also compute, from a purely algebraic perspective, the effects registered by a left universe observer due to a unitary fluctuation on the right universe of the traversable wormhole, and reproduce a result from an earlier computation by Maldacena, Stanford and Yang \cite{Maldacena:2017axo}.}

\begin{document}
\maketitle
\flushbottom

\section{Introduction}
\label{sec:intro}

AdS/CFT has been a remarkable tool for understanding quantum gravity; and is a non perturbatively well defined setup to look into the fine details of the quantum nature of gravity. An interesting instance is the conjecture of Maldacena \cite{Maldacena:2001kr}, following the work of Israel \cite{Israel:1976ur}, that the thermally entangled state of two boundary CFT's is a non perturbative description of the maximally extended AdS-Schwarzchild black hole. It is therefore natural to probe deeply the mathematical structures of the boundary CFT to clarify the properties of the bulk gravity theory. To this end, in recent years, a quantum information theoretic/quantum computation and operator algebraic discussions of the boundary have become important. It was noticed that \cite{Leutheusser:2021frk,Leutheusser:2022bgi} the extreme large $N$ ($N=\infty$) limit of each of the thermally entangled CFT's is given a type III$_{1}$ von Neumann algebra, as would be expected from the bulk dual, while the addition of $1/N$ corrections changed the algebra to type II$_{\infty}$ factor \cite{ Witten:2021unn}. The same can be said about boundary CFT subregions, which again at the extreme large $N$ limit are expected to be type III$_{1}$ either from the bulk or the boundary perspective \cite{Bahiru:2022mwh}. If one would rather consider the micro-canonically entangled CFT's, the extreme large $N$ limit would itself be a type II$_{\infty}$ algebra, where the bulk theory even at the extreme limit contains a gravitational degree of freedom not present in the naive QFT on a curved spacetime limit \cite{Chandrasekaran:2022eqq,Bahiru:2023ify}\footnote{For the microcanonical state the variance in the energy is $O(1)$, therefore a subtraction of the one point function is enough to have a well defined one sided Hamiltonian in the large N limit. For this reason, the Hamiltonian produces an $O(1)$ transformation in the large N Hilbert space and operators (as opposed to $O(1/N)$ transformation as in the case for the canonical ensemble). The conjugate mode to the one sided Hamiltonian has a nice bulk interpretation as a non local mode which produces time shifts between the left and right boundary of the bulk Cauchy slice. Even though one is looking at the $N =\infty$ theory, states that differ by the time shift mode correspond to different semiclassical spacetimes. Thus this mode is absent in the naive QFT on a curved spacetime limit however it emerges in the large N limit of the microcanonical ensemble.}, it was also found that the presence of positive energy shock waves would lead to a free product von Neumann algebras. Many more interesting examples have been worked out some of which include\cite{Chandrasekaran:2022cip,Kudler-Flam:2023qfl,Kudler-Flam:2024psh,Penington:2023dql,Witten:2023xze,Penington:2024sum,Aguilar-Gutierrez:2023odp,Bahiru:2023ify,Jensen:2024dnl,deBoer:2025rxx,Leutheusser:2022bgi,Leutheusser:2024yvf,Leutheusser:2025zvp,Gesteau:2024rpt}.

This work follows the same spirit in that the goal is understanding the boundary theory better to learn more about the bulk theory, with a particular focus on the bulk eternal black hole being a traversable one. A traversable wormhole as a relevant double trace deformation of the boundary CFT's was discussed in \cite{Gao:2016bin}. In addition to leading to the first example of traversable wormhole with a UV complete description, the Gao-Jafferis-Wall protocol elaborated on the deep connection of the bulk and the boundary in that a boundary perturbation on first CFT really does travel through the bulk wormhole spacetime and emerges in the second universe to its boundary \cite{Maldacena:2017axo}. Interesting connections with an earlier work of the butterfly effect and shock waves in AdS/CFT was also found \cite{Shenker:2013pqa}. We have provided in this article an algebraic description of this story by considering a novel large $N$ limit where we take the large $N$ limit in the presence of a certain operator, an element from an algebra of operators closely related to algebra at infinity discussed in \cite{Lechner:2021fvy}. This operator will produce the necessary negative energy shock wave to make the deep bulk a traversable wormhole. The presence of this operator modifies the expectation values of operators on the left and right CFT's leading to a new large $N$ Hilbert space for the traversable wormhole, following the standard GNS construction. When the closure of the algebra of operators in the CFT 1 and 2 is done with respect to this Hilbert space, one gets the algebra of operators for the traversable wormholes.    

In addition, we will reproduce a result for the back reaction of the double trace deformation computed in \cite{Maldacena:2017axo} which involves computing the two point function in the traversable wormhole background. This two point function is important in the construction of the Hilbert space for traversable wormholes as it shows the distinction of this new Hilbert space from the standard GNS Hilbert space of eternal black holes. In \cite{Maldacena:2017axo} the authors used an input from an earlier computation of gravitational scattering to arrive at the result. However in this paper, without making use of the behavior of the gravitational scattering, we compute the back-reaction using solely algebraic tools and starting from finite $N$ and taking the large $N$ limit in the presence of the negative energy shock wave operator. This suggests at least this particular result of the gravitational scattering was already encoded in the algebraic properties of the large $N$ limit. In the computation, we have assumed that the large $N$ limit of the thermofield state is cyclic and separating for a semi infinite time band subalgebra of the full large $N$ algebra of each CFT, fig. \ref{fig.hsmi}. This is necessary so that the full large $N$ algebra and the semi infinite time band subalgebra can be taken to satisfy the half sided modular inclusions. An earlier discussion on traversable wormhole from algebraic perturbation theory perspective can also be found in \cite{AliAhmad:2025ukh}.   

In the next section we will start with some comment on traversability and the concept of sub-systems in algebraic quantum field theory, which was made precise by a property called the split property and its relation to the conjecture in \cite{Kawamoto:2025oko} on the signature for traverasbility of wormholes. Then in section \ref{traver worm}, we review the Gao-Jafferis-Wall traversable wormhole and its connection to shock waves in eternal black hole and discuss the computation of the effects of a unitary perturbation of one boundary on the second boundary. In section \ref{alg tra worm}, we present the set up and show the computation that reproduces the result from the previous section. Finally in the last section, we discuss the Hilbert space and algebra of operators for traversable wormholes; before concluding with a discussion of the result and some possible future follow up works. The Appendix also includes some of the relevant discussion of the mathematical background and motivation for this paper.      

\section{Traversability}\label{sec:trave}

The main goal of this section is to introduce the relevant objects and computations that will be used in the following sections, and in particular emphasize the role of the two point function in the eternal black hole and traversable wormhole background, which is an essential element in this paper. In addition, we will also comment on the recent conjecture \cite{Kawamoto:2025oko} on the signature for traversablity of wormhole geometries. 

A useful way to analyze traversablity is a boundary two point function involving the single trace operators of the two entangled CFTs. To be more precise, we begin with defining the thermofield double state in the semiclassical limit, $\ket{\Psi}$, as the large $N$ (or the appropriate\footnote{In general one takes the limit where the degrees of freedom of the CFT are large, that is the central charge$(c)$ (typically the central charge that appears in the OPE of two stress energy tensors) of the CFT is taken to infinity. For $\mathcal{N}=4$ super Yang Mills in $4d$ with $SU(N)$ gauge group, we have $c \sim N^{2}$.} limit of the boundary CFT) of the thermally entangled state of the two CFTs at temperatures above the Hawking-Page temperature by

\begin{equation}\label{deftfd}
    \bra{\Psi}O\ket{\Psi} = \lim_{N\rightarrow \infty} \bra{TFD}O\ket{TFD}_{N},
\end{equation}

for any bounded functions of the appropriately normalized single trace operators, i.e, where the one point function is subtracted\footnote{As opposed to the case where one just divides the single trace operators by $N$ leading to a Poisson algebra of the operators, which is dual to the classical phase space of the bulk gravitational theory \cite{Witten:2023xze}.}, $O$, and a state in the finite $N$ CFT

\begin{equation}
    \Ket{TFD} = \frac{1}{\sqrt{Z}} \sum_{i} e^{-\frac{\beta E_{i}}{2}} \ket{E_{i}}_{1}\ket{E_{i}}_{2}
\end{equation}
 where $Z$ is such that $ \Ket{TFD}$ is normalized.

 The two point function $\bra{\Psi}O_{1}O_{2}\ket{\Psi}$ is in general non-trivial even when the two CFTs are non interacting. This is solely a signature of the entanglement present in the underlying state between the two systems. Since we construct the semiclassical Hilbert space by taking the closure of a set constructed by acting with bounded functions of $O$'s on $\ket{\Psi}$, any state has non-trivial two point functions in general. 

 This, however, is not implying a causal connection between the two systems and there is no 'information transfer' between the two CFTs. Importantly the algebra of operators associated with the two CFTs commute, which is the statement of microcausality. A stricter condition is expected to be satisfied for free theories, whose mass spectrum grows fast enough, thermodynamic theories or more generally quantum field theories considered, in the early days of algebraic QFT, to be `physical'. In particular, the algebras of operators associated with the two CFTs (the more precise statement is not about algebras of CFT operators for all times rather there should be an arbitrarily big(for far future) and small(for the far past) upper and lower bounds respectively), $O_{1} \in \mathcal{A}_{1}$  and $O_{2} \in \mathcal{A}_{2}$, \emph{or} measurements done on the two systems are statistically independent. Given the definition \eqref{deftfd}, we can think of a state as a collection of expectation values for all operators in $\mathcal{A}_{i}$\footnote{In the standard discussion in algebraic QFT, one actually starts by defining states as positive normalized linear form on the algebra of operators then the Hilbert space and state vectors in the Hilbert space follow from the so called GNS construction \cite{Haag:1996hvx}.} i.e, a linear functional, $\varphi$ on the algebra of operators with the appropriate conditions that follow from a normalized state vector like $\ket{\Psi}$, which includes reality, positivity and normalizability\footnote{More precisely, we have $\varphi(O^{\star}) =\overline{\varphi(O)}$, $\varphi(O^{\star} O) \geq 0$ and  $||\varphi||=1$ where $||\varphi|| =$ sup$_{O \in \mathcal{A}} \frac{|\varphi(O)|}{||O||}$.}. 
 
 These functionals encode expectation values of repeated measurements for any of the operators, and statistical independence for measurements of operators in $\mathcal{A}_{i}$ is a statement on a combined algebra of operators $(\mathcal{A}_{1} \cup \mathcal{A}_{2})^{''} := \mathcal{A}$, that any two normal states $\varphi_{1}$ and $\varphi_{2}$ on $\mathcal{A}_{1}$ and $\mathcal{A}_{2}$ can be extended to a normal state\footnote{These are of course normal states with respect to the appropriate cyclic and separating state. For our discussion above, it is $\ket{\Psi}$; for some of the discussion that follows concerning Minkowski spacetime, it is the Minkowski vacuum $\Ket{\Omega}$.} $\varphi$ on 
$\mathcal{A}$ such that,

\begin{equation}\label{splst}
    \varphi(O_{1}O_{2}) = \varphi_{1}(O_{1})\varphi_{2}(O_{2}).
\end{equation}

This statement, usually known as the split property, follows from a condition called the nuclearity condition. This involves several requirements including the requirement on how big two point functions like $\bra{\Psi}O_{1}O_{2}\ket{\Psi}$ can be, so as to identify states that can be thought of as `essentially localized' \cite{Haag1965}.

The reason for this possibility is the fact that for larger and larger spacelike distances between the two operators in our correlation function, the correlation function, which as was mentioned earlier is due to the underlying entanglement of the vacuum $\ket{\Omega}$, decays. Therefore, if one is interested in creating a state localized in region (more precisely a causal diamond) $\mathcal{R}_{1}$, then one uses an operator localized in $\mathcal{R}_{1}$, say $a_{1}$. However, if one intends to create the state by acting with an operator $a_{2}$ localized far away in a spacelike separated region, $\mathcal{R}_{2}$, then the norm of this vector will be extremely small, since correlations between far away regions is small. Making use of this fact, one can restrict to states that are `essentially localized' in $\mathcal{R}_{2}$ by looking at states that are created by the action of operators in $\mathcal{R}_{2}$,
\begin{equation}
   \ket{\Phi} = A\ket{\Omega}\;\text{, with } A\in \mathcal{A(R}_{2})
\end{equation}
and defining 
\begin{equation}
    c_{A} = \frac{||A||}{|A\ket{\Omega}|}
\end{equation}

and requiring to only look at states with $c_{A}<c$, for some fixed number $c$, therefore excluding states with very small norm corresponding to far-away-localized excitations. This was the original motivation of \cite{Haag1965}, but later stronger conditions were discussed in several follow up works \cite{Buchholz:1986dy,Buchholz:1973bk,Doplicher:1984zz,Buchholz:1986bg,Buchholz:1989zz,Buchholz:1989eb}. For more recent discussions including curved spacetime backgrounds, check \cite{Verch:1993fs,Fewster:2015acv,fewster2016split}.   

Therefore this condition by far does not require the two point function to vanish however it can not be arbitrarily big, in particular if one modifies the system such that a singularity arises for this two point function\footnote{To be precise, this is the two point function for the vacuum or for the relevant cyclic separating state in general cases.}, one can not really use the above argument to restrict to `essentially localized' states and find a split state $\varphi$ with the property \eqref{splst}, since $c_{A}= 0$. Therefore one can not say $\mathcal{A}_{1}$ and $\mathcal{A}_{2}$ are statistically independent. 

In the situations where one can explicitly check microcausality, of course the failure of microcausality is a stronger condition than the failure of split property. However, it was conjectured in \cite{Kawamoto:2025oko} that a singularity in the `left - right' correlation function was signal of traversablity. What we have argued here is that a singularity in the `left - right' two point function implies the failure of the split property for the two regions, which is a step towards showing a causal connection.
%In other words there is causal connection between the two systems and information transfer is possible. 

In AdS/CFT, the state $\ket{\Psi}$ is dual to the Hartle Hawking vacuum on the background of eternal black hole in asymptotically AdS spacetime. The single trace operators are identified with the boundary limit of bulk field operators with the appropriate prefactor to cancel out the decaying factor in the coordinate orthogonal to the conformal boundary, 

\begin{equation}\label{bulktoboun}
    \bra{\Psi}O_{1}O_{2}\ket{\Psi} = \lim_{\epsilon \rightarrow 0} \epsilon^{-2\Delta} \bra{HH}\phi_{1,\epsilon}\phi_{2,\epsilon}\ket{HH}
\end{equation}
where $\ket{HH}$ is the Hartle Hawking vacuum constructed by a Euclidean path integral on a hemisphere, i.e, half of Euclidean eternal black hole geometry. $\phi_{i}$ are bulk fields dual to the boundary operators $O_{i}$ and $\Delta$ is the conformal dimension of the $O_{i}$'s. 

The bulk two point function is the Green's function \cite{Balasubramanian:1999zv} for the free theory in the bulk\footnote{The interactions in the bulk supergravity theory are controlled by $1/c$ and there for in the large $c$ limit, the bulk becomes free. This is particularly true in the extreme large N limit. The boundary description of this theory in the bulk is given by generalized free field theory.} in the eternal black hole background, therefore one has,

\begin{equation}\label{pathequ}
    \bra{HH}\phi_{1}\phi_{2}\ket{HH} = \int D\mathcal{P} e^{-\Delta L(\mathcal{P})} 
\end{equation}
where the integral is over paths connecting the locations of the two bulk fields and $D\mathcal{P}$ is a measure for such curves with the appropriate Faddeev Popov factor, which follows from the gauge fixing along the curves \cite{Louko:2000tp}. $L(\mathcal{P})$ is the regularized length of each path $\mathcal{P}$ and it is real and positive for spacelike curves while imaginary for time like curves. 

In the large conformal dimension limit, one can use the saddle point approximation to get,

\begin{equation}\label{twopoigeo}
    \bra{HH}\phi_{1}\phi_{2}\ket{HH} \sim \sum_{\text{saddles}} e^{-\Delta L}
\end{equation}

The simplest example to do this computation and observe the appearance of the singularity is the BTZ black hole. The BTZ black hole spacetime can thought as a compactification of AdS$_{3}$ along some Killing vector field therefore the dominant contribution to the two point function coming from the geodesic length can be seen from the geodesic length in pure AdS$_{2+1}$. The metric is given by,

\begin{equation}
    ds^{2}= - \frac{r^{2}-R^{2}}{l^{2}} dt^{2}+\frac{l^{2}}{r^{2}-R^{2}} dr^{2} +r^{2}d\varphi^{2}
\end{equation}
where $\varphi \simeq \varphi +2\pi$, and $R$ is the horizon radius while $l$ is the AdS size. 

The geodesic length between two points in the embedding space $(T_{1},T_{2},X_{1},X_{2})$ and $(T^{'}_{1},T^{'}_{2},X^{'}_{1},X^{'}_{2})$ in pure AdS$_{2+1}$ is given as 

\begin{equation}
    \text{cosh}\frac{d}{l} = T_{1}T^{'}_{1}+T_{2}T^{'}_{2}-X_{1}X^{'}_{1}-X_{2}X^{'}_{2}
\end{equation}
with the embedding coordinates 

\begin{align*}
   T_{1} &= \frac{1}{R}\sqrt{r^{2}-R^{2}} \text{ sinh}\frac{Rt}{l^{2}} \\
   T_{2} &= \frac{r}{R} \text{ cosh}\frac{R\varphi}{l}\\
   X_{1} &= \frac{1}{R}\sqrt{r^{2}-R^{2}} \text{ cosh}\frac{Rt}{l^{2}} \\
   X_{2} &= \frac{r}{R} \text{ sinh}\frac{R\varphi}{l}
\end{align*}

we recover the BTZ black hole. Therefore, one can easily check the geodesic distance between points $(r,t_{1},\varphi)$ and $(r,t_{2},\varphi)$ for large $r$, i.e, as we get closer to the boundary. We get

\begin{equation}
    \frac{d}{l} = 2 \text{ log}\frac{2r}{R} + 2 \text{ log}[\text{ cosh}(\frac{R}{2l^{2}}(t_{2}+t_{1}))]
\end{equation}

However as we will see in section \ref{shock trav}, once we modify the spacetime to get a traversable wormhole, the geodesic distance will be 
\begin{equation}
    \frac{d_{\text{trav}}}{l} = 2\text{ log}\frac{2r}{R} + 2 \text{ log}\left[ \text{ cosh}(\frac{R (t_{1}+t_{2})}{ 2l^2}) + \frac{\alpha}{2} e^{-\frac{\pi}{\beta} (-t_{1}+t_{2})} \right]
\end{equation}
depending on a certain parameter $\alpha$. To recover the boundary two point function, one has to regularize the geodesic distance, which would remove the first term. Then the boundary two point function is given by

\begin{equation}
    \langle O_{1}(t_{1}=0,\varphi)O_{2}(t_{2}=0,\varphi)\rangle = e^{-2ml \text{ log}(1+\frac{\alpha}{2})} = \left(\frac{1}{1+\frac{\alpha}{2}}\right)^{2ml}
\end{equation}

Therefore as $\alpha \rightarrow -2$, a singularity emerges signaling a causal connection between the two boundaries. Several more examples can be found in \cite{Kawamoto:2025oko}.

\section{Traversable wormholes}\label{traver worm}

\subsection{Gao-Jafferis-Wall traversable wormholes}

Gao, Jafferis and Wall proposed the first traversable wormhole solution that can be embedded in a UV complete gravitational theory, \cite{Gao:2016bin}. Topological censorship prohibits the existence of traversable wormholes if the null energy condition or even the average null energy condition (ANEC) is satisfied \cite{Friedman:1993ty, Witten:2019qhl}. Even though in quantum field theory all the point-wise energy conditions are violated essentially as a result of the Reeh Schlieder theorem \cite{Witten:2018zxz}, ANEC is still obeyed in most cases\cite{Wald:1991xn,Penrose:1993ud,Graham:2007va,Faulkner:2016mzt,Hartman:2016lgu,Kontou:2015yha}. They proposed to produce a violation of ANEC, and so construct a traversable wormhole, by coupling the two CFTs that are already thermally entangled, by introducing a time dependent interaction. The interaction is a relevant double trace deformation of the CFTs, therefore deep in the IR it modifies the geometry making it traversable, while the modification is negligible in the UV, i.e, no modification to the geometry close to the boundary. Since the thermally entangled state between the two CFTs is a UV complete description of the eternal black hole \cite{Maldacena:2001kr}, a relevant deformation is still a consistent solution in AdS/CFT.  

In more precise terms, the double trace deformation is such that it will modify the boundary conditions for a bulk scalar field with the appropriate dimensions\footnote{This is so that the double trace operator is a relevant operator as desired.}. After analytical and numerical computations, for the appropriate sign of $h$, they showed that such deformation will lead to a contribution to the stress energy tensor that violates ANEC. More explicitly, the deformation is the addition of the following term to the total Hamiltonian of the CFTs,
\begin{equation}\label{pert. hamil}
    \delta H(t) = - \int d\bar{x} h(t,\bar{x}) O_{1}(-t,\bar{x}) O_{2}(t,\bar{x})
\end{equation}
where $h(t,\bar{x}) = 0$ for $t<t_{0}$.

To see how the violation of ANEC leads to a traversable wormhole geometry, let's follow a light ray along the `left-to-right horizon'. If in the Kruskal coordinates for the eternal black hole $u$ and $v$ are the two null coordinates. The coordinates $u=v=0$ is the bifrication point and the horizon at $u =0$ can be parametrized by $u$ and approaches the left boundary as $t \rightarrow -\infty$ and approaches the right boundary as $t \rightarrow \infty$. 

After the deformation is introduced, there will be a perturbation, $h_{\mu\nu}$, to the original metric and the coordinate for the light ray that originated in the past horizon is given by  
\begin{equation}
    u(v)=-\frac{1}{2g_{vu}(u=0)} \int_{-\infty}^{v} dv\;h_{vv} 
\end{equation}
where $g_{vu} <0 $ is the components of the original metric. On the other hand,
\begin{equation}
    \int dv \; h_{vv} \propto 8\pi G_{N} \int dv T_{vv}
\end{equation}
with a positive proportionality constant for spacetime dimensions $d\geq 3$. Therefore, if ANEC is violated i.e,
\begin{equation}
    \int_{-\infty}^{\infty} dv \;T_{vv} <0,
\end{equation}
then $u(\infty)<0$, which means the light ray will reach the right boundary. 

The time dependence is important here since in the undeformed case, $\int dv\;T_{vv}$ annihilates the thermofield double state where the contribution from $v<0$ exactly cancels out the contribution from $v>0$. The additional time dependent interaction is introducing a discrepancy between these two contributions. One can also think of this prescription as a quantum teleportation protocol in the ER=EPR context where part of the non local interaction is modeled by the classical information necessary in quantum teleportation \cite{Maldacena:2017axo}.

\subsection{Shock waves and traversable wormholes}\label{shock trav}

Note that we have explored the geometry of the traversable wormhole above, following a light ray that originated in the far past and emerged in the right universe in the far future. Below, we discuss a closely related but alternative perspective of the same system where it is the operators of the double trace deformation that are `put' in the far past and far future, using the symmetry of the eternal black hole. The double trace deformation can be understood as a creation of an $O(1)$ energy particle in the far past and measuring it in the far future. It is useful to arrive at this perspective from a slightly different situation where we just send a small perturbation from the region close to the boundary into the black hole. This perturbation can be assumed to emerge from the white hole of the geometry by time evolving it in to the past with a time independent Hamiltonian. If it emerged from the boundary at far enough time, due to the coordinate transformation between the Kruskal coordinates near a black hole horizon and Schwarzchild coordinates at infinity, small perturbations at infinity however will be blue shifted shockwaves as they get close to horizon, producing a large back-reaction in the deep bulk. This effect was discussed in \cite{Shenker:2013pqa} and this back reaction was considered for the case of eternal black hole in asymptotically AdS spacetimes and its boundary dual thermofield double state. 

A small perturbation of local energy $E_{+}$ sent from the far past $t_{1} = - t_{0}$ at the left boundary of the eternal black hole will have a blue shifted energy $E \sim E_{+} e^{\frac{2\pi t_{0}}{\beta}}$ as it gets close to the black hole horizon. When $E$ gets big enough, that is $t_{0}$ is taken to be big, the perturbation will take an almost null path and one will have to consider its back reaction on the eternal black hole geometry. In particular, following an earlier computation for the asymptotically flat case \cite{Dray:1984ha,Dray:1985yt}, the back reacted geometry is a gluing of two black hole spacetimes along the shock wave localized at $u = e^{-\frac{2\pi t_{0}}{\beta}}$, with ADM masses $M$ and $M+E_{+}$, where $M$ is the ADM mass of the unperturbed black hole. 

\begin{figure}

\begin{tikzpicture}[scale=3.6]

  \def\R{0.08} % size lightcone
  \def\Nlines{3} % number of world lines (at constant r/t)
  \pgfmathsetmacro\ta{1/sin(90*1/(\Nlines+1))} % constant r/t value 1
  \pgfmathsetmacro\tb{sin(90*2/(\Nlines+1))}   % constant r/t value 2
  \pgfmathsetmacro\tc{1/sin(90*2/(\Nlines+1))} % constant r/t value 3
  \pgfmathsetmacro\td{sin(90*1/(\Nlines+1))}   % constant r/t value 4
  \coordinate (-O) at (-1, 0); % center III: origin (r,t) = (0,0)
  \coordinate (-S) at (-0.5,-1); % south III: t=-infty, i-
  \coordinate (-N) at (-1, 1); % north III: t=+infty, i+
  \coordinate (-W) at (-0.75, 0); % east III:  r=-infty, i0
  %\coordinate (L) at (0.55, 0.55); % N, 
  %\coordinate (M) at (1, 0.1); % N, 
  \coordinate (-E) at ( 0.15, -0.133); % west III:  r=+infty, i0
   \coordinate (C) at ( 0.25, 0); % west III:  r=+infty, i0
  \coordinate (O)  at ( 1, 0); % center I: origin (r,t) = (0,0)
  \coordinate (S)  at ( 1.5,-1); % south I: t=-infty, i-
  \coordinate (N)  at ( 1, 1); % north I: t=+infty, i+
  %\coordinate (-L) at (-0.55, -0.55); % JNJ,
  % \coordinate (-M) at (-1, -0.1); % N,
  \coordinate (E)  at ( 1.25, 0); % east I:  r=-infty, i0
  \coordinate (W)  at ( 0.35, 0.133); % west I:  r=+infty, i0
  \coordinate (B)  at ( 0,-1); % singularity bottom
  \coordinate (T)  at ( 0, 1); % singularity top
  \coordinate (X0) at ({asin(sqrt((\ta^2-1)/(\ta^2-\tb^2)))/90},
                       {-acos(\ta*sqrt((1-\tb^2)/(\ta^2-\tb^2)))/90}); % particle 1
  \coordinate (X1) at ({asin(sqrt((\tc^2-1)/(\tc^2-\td^2)))/90},
                       {acos(\tc*sqrt((1-\td^2)/(\tc^2-\td^2)))/90}); % particle 2
  \coordinate (X2) at (45:0.87); % particle falling in BH horizon
  \coordinate (X3) at (0.60,1.05); % particle falling in BH singularity
  
  \begin{scope}
    
    % CLIP to fill inside zigzag lines
    \clip[decorate,decoration={zigzag,amplitude=2,segment length=6.17}]
      (S) -- (-S) --++ (-1.1,-0.1) |-++ (4.2,2.2) |- cycle;
    \clip[decorate,decoration={zigzag,amplitude=2,segment length=6.17}]
      (-N) -- (N) --++ (1.1,0.1) |-++ (-4.2,-2.2) |- cycle;
    
    % REGIONS FILLS
    \fill[mylightpurple] (-N) |-++ (2,0.1) -- (N) -- (-S) -- (S) -- cycle;
    \fill[mylightpurple] (-S) |-++ (2,-0.1) -- (S) -- (-N) -- (N) -- cycle;
    \fill[mylightpurple] (N) -- (-E) -- (-N)  -- cycle;
     \fill[mylightpurple] (S) -- (W) -- (-S)  -- cycle;
    \fill[mylightblue] (-N) -- (-E) -- (-S) -- (-W) -- cycle;
    \fill[mylightblue] (N) -- (E) -- (S) -- (W) -- cycle;
   
  \end{scope}
  
  % BOUNDARIES
  \draw[singularity] (-N) -- node[above] {} 
  (N);
  \draw[singularity] (S) -- node[below] {} (-S);
  \path (S) -- (W) node[mydarkblue,pos=0.50,below=-2.5,rotate=-45,scale=0.85]
 {\contour{mylightpurple}{}};
 % \path (L) -- (M) node[mydarkblue,pos=0.40,below=-2.5,rotate=-45,scale=0.85]
   {\contour{mylightpurple}{}};
  \path (W) -- (N) node[mydarkblue,pos=0.32,above=-2.5,rotate=45,scale=0.85]
    {\contour{mylightpurple}{}};
  \draw[thick,mydarkblue] (-N) -- (-E) -- (-S) -- (-W) -- cycle;
  %\draw[thick,mydarkblue] (N) -- (L) -- (M) -- cycle;
  %\draw[thick,mydarkblue] (-S) -- (-L) -- (-M) -- cycle;
  \draw[thick,mydarkgreen] (N) -- (W) -- (-S);
  \draw[thick,mydarkblue] (N) -- (E) -- (S) -- (W);
  
  % REGIONS
  %\node[fill=mylightblue,inner sep=2] at (-O) {};
  %\node[fill=mylightblue,inner sep=2] at (O) {};
  %\node[fill=mylightpurple,inner sep=2] at (0,0.64) {};
  %\node[fill=mylightpurple,inner sep=2] at (0,-0.64) {};
  
  % INFINITY LABELS
  \node[above=1,left=1,mydarkblue] at (-1,0.5) {$\mathcal{M^{'}}$};
  %\node[above=1,right=1,mydarkblue] at (1,0) {$t=0$};
  \node[above=1,right=1,mydarkblue] at (1.5,-0.5) {$\mathcal{M}$};
   %\node[above=0.5,right=1,mydarkblue] at (1,0.5) {$\mathcal{N}$};
    %\node[left=1,below=0.5,mydarkblue] at (-1.2,-0.5) {$J\mathcal{N}J$};
    \node[left=1,below=0.5,mydarkpurple] at (-0.75,-1) {CFT$-1$};
    \node[above=1,right=1,mydarkpurple] at (1.5,-1.05) {CFT$-2$};
  %\node[right=1,below=1,mydarkpurple] at (-S) {$i^-$};
  %\node[right=1,above=1,mydarkpurple] at (-N) {$i^+$};
  %\node[right=1,below=1,mydarkpurple] at (S) {$i^-$};
  %\node[right=1,above=1,mydarkpurple] at (N) {$i^+$};
  %\node[mydarkblue,below left=-1] at (-1.5,-0.5) {$\calI^-$};
  %\node[mydarkblue,above left=-1] at (-1.5,0.5) {$\calI^+$};
  %\node[mydarkblue,above right=-1] at (1.5,0.5) {$\calI^+$};
  %\node[mydarkblue,below right=-1] at (1.5,-0.5) {$\calI^-$};
  
  % LIGHT CONE FRONT
  %\conefront{X0};
  %\conefront{X1};
  %\conefront{X2};
  
  % ESCAPING PHOTONS
  %\draw[photon] (X0) ++ (45:0.1) --++ (45:0.3);
  %\draw[photon] (X1) ++ (45:0.1) --++ (45:0.3);
  
\end{tikzpicture}

\begin{tikzpicture}[scale=3.6]

  \def\R{0.08} % size lightcone
  \def\Nlines{3} % number of world lines (at constant r/t)
  \pgfmathsetmacro\ta{1/sin(90*1/(\Nlines+1))} % constant r/t value 1
  \pgfmathsetmacro\tb{sin(90*2/(\Nlines+1))}   % constant r/t value 2
  \pgfmathsetmacro\tc{1/sin(90*2/(\Nlines+1))} % constant r/t value 3
  \pgfmathsetmacro\td{sin(90*1/(\Nlines+1))}   % constant r/t value 4
  \coordinate (-O) at (-1, 0); % center III: origin (r,t) = (0,0)
  \coordinate (-S) at (-1.5,-1); % south III: t=-infty, i-
  \coordinate (-N) at (-1, 1); % north III: t=+infty, i+
  \coordinate (-W) at (-1.25, 0); % east III:  r=-infty, i0
  %\coordinate (L) at (0.55, 0.55); % N, 
  %\coordinate (M) at (1, 0.1); % N, 
  \coordinate (-E) at ( -0.15, 0.08); % west III:  r=+infty, i0
  \coordinate (O)  at ( 1, 0); % center I: origin (r,t) = (0,0)
  \coordinate (S)  at ( 0.5,-1); % south I: t=-infty, i-
  \coordinate (N)  at ( 1, 1); % north I: t=+infty, i+
 % \coordinate (-L) at (-0.55, -0.55); % JNJ,
  % \coordinate (-M) at (-1, -0.1); % N,
  \coordinate (E)  at ( 0.75, 0); % east I:  r=-infty, i0
  \coordinate (W)  at ( -0.35, -0.08); % west I:  r=+infty, i0
  \coordinate (B)  at ( 0,-1); % singularity bottom
  \coordinate (T)  at ( 0, 1); % singularity top
  \coordinate (X0) at ({asin(sqrt((\ta^2-1)/(\ta^2-\tb^2)))/90},
                       {-acos(\ta*sqrt((1-\tb^2)/(\ta^2-\tb^2)))/90}); % particle 1
  \coordinate (X1) at ({asin(sqrt((\tc^2-1)/(\tc^2-\td^2)))/90},
                       {acos(\tc*sqrt((1-\td^2)/(\tc^2-\td^2)))/90}); % particle 2
  \coordinate (X2) at (45:0.87); % particle falling in BH horizon
  \coordinate (X3) at (0.60,1.05); % particle falling in BH singularity
  
  \begin{scope}
    
    % CLIP to fill inside zigzag lines
    \clip[decorate,decoration={zigzag,amplitude=2,segment length=6.17}]
      (S) -- (-S) --++ (-1.1,-0.1) |-++ (4.2,2.2) |- cycle;
    \clip[decorate,decoration={zigzag,amplitude=2,segment length=6.17}]
      (-N) -- (N) --++ (1.1,0.1) |-++ (-4.2,-2.2) |- cycle;
    
    % REGIONS FILLS
    \fill[mylightpurple] (-N) |-++ (2,0.1) -- (N) -- (-S) -- (S) -- cycle;
    \fill[mylightpurple] (-S) |-++ (2,-0.1) -- (S) -- (-N) -- (N) -- cycle;
    
    \fill[mylightblue] (-N) -- (-E) -- (-S) -- (-W) -- cycle;
    \fill[mylightblue] (N) -- (E) -- (S) -- (W) -- cycle;
     \end{scope}

  % BOUNDARIES
  \draw[singularity] (-N) -- node[above] {} 
  (N);
  \draw[singularity] (S) -- node[below] {} (-S);
  \path (S) -- (W) node[mydarkblue,pos=0.50,below=-2.5,rotate=-45,scale=0.85]
 {\contour{mylightpurple}{}};
 % \path (L) -- (M) node[mydarkblue,pos=0.40,below=-2.5,rotate=-45,scale=0.85]
   {\contour{mylightpurple}{}};
  \path (W) -- (N) node[mydarkblue,pos=0.32,above=-2.5,rotate=45,scale=0.85]
    {\contour{mylightpurple}{}};
  \draw[thick,mydarkblue] (-N) -- (-E) -- (-S) -- (-W) -- cycle;
  \draw[thick,mydarkred] (N) -- (W) -- (-S);
  %\draw[thick,mydarkblue] (N) -- (L) -- (M) -- cycle;
  %\draw[thick,mydarkblue] (-S) -- (-L) -- (-M) -- cycle;
  \draw[thick,mydarkblue] (N) -- (E) -- (S) -- (W);
  
  % REGIONS
  %\node[fill=mylightblue,inner sep=2] at (-O) {};
  %\node[fill=mylightblue,inner sep=2] at (O) {};
  %\node[fill=mylightpurple,inner sep=2] at (0,0.64) {};
  %\node[fill=mylightpurple,inner sep=2] at (0,-0.64) {};
  
  % INFINITY LABELS
  \node[above=1,left=1,mydarkblue] at (-1.2,0.5) {$\mathcal{M^{'}}$};
  %\node[above=1,right=1,mydarkblue] at (1,0) {$t=0$};
  \node[above=1,right=1,mydarkblue] at (0.75,-0.5) {$\mathcal{M}$};
 %  \node[above=0.5,right=1,mydarkblue] at (1,0.5) {$\mathcal{N}$};
%    \node[left=1,below=0.5,mydarkblue] at (-1.2,-0.5) {$J\mathcal{N}J$};
    \node[left=1,below=0.5,mydarkpurple] at (-1.65,-1) {CFT$-1$};
    \node[above=1,right=1,mydarkpurple] at (0.55,-1.05) {CFT$-2$};
  %\node[right=1,below=1,mydarkpurple] at (-S) {$i^-$};
  %\node[right=1,above=1,mydarkpurple] at (-N) {$i^+$};
  %\node[right=1,below=1,mydarkpurple] at (S) {$i^-$};
  %\node[right=1,above=1,mydarkpurple] at (N) {$i^+$};
  %\node[mydarkblue,below left=-1] at (-1.5,-0.5) {$\calI^-$};
  %\node[mydarkblue,above left=-1] at (-1.5,0.5) {$\calI^+$};
  %\node[mydarkblue,above right=-1] at (1.5,0.5) {$\calI^+$};
  %\node[mydarkblue,below right=-1] at (1.5,-0.5) {$\calI^-$};
  
  % LIGHT CONE FRONT
  %\conefront{X0};
  %\conefront{X1};
  %\conefront{X2};
  
  % ESCAPING PHOTONS
  %\draw[photon] (X0) ++ (45:0.1) --++ (45:0.3);
  %\draw[photon] (X1) ++ (45:0.1) --++ (45:0.3);
  
\end{tikzpicture}

\caption{\emph{(above)} is the Penrose diagram of a wormhole in the presence of a shock wave with positive energy very close to the horizon (shown is green), i.e, sent from the very past in CFT$-1$. The future horizon of the left black hole and the past horizon of the right black hole will 'miss' each other because of the time advance geodesics receive in the presence of the shock wave. However \emph{(below)}, if the energy of the shock wave is negative (shown in dark red), the  infalling geodesics will translate backwards along the trajectory of the shock wave and can escape to the left black hole's asymptotic universe. }\label{fig.}
\end{figure}
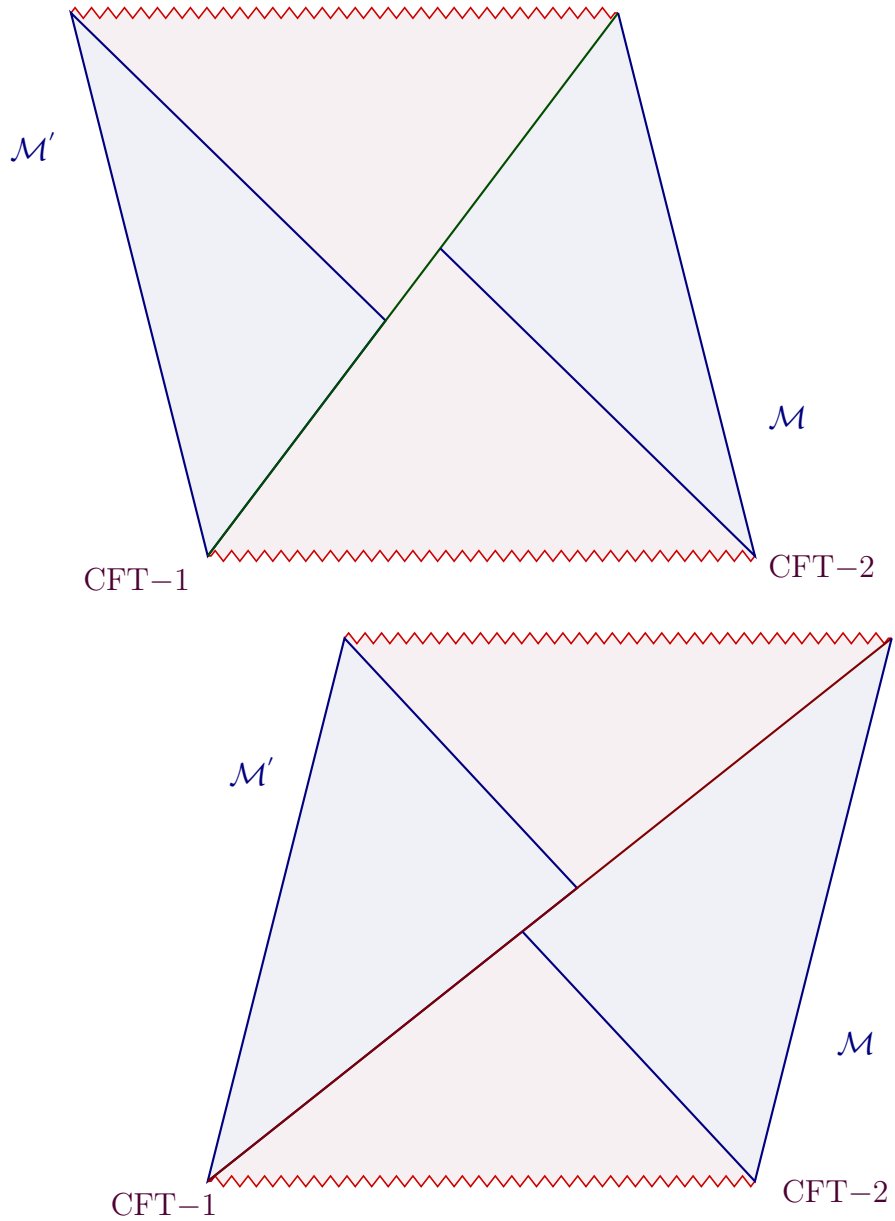

Consider for instance the BTZ black hole,

\begin{equation}
    ds^{2}= \frac{-2 \beta R_{2}/\pi \; dudv + R_{2}^{2}(1-uv)^{2} d\phi^{2}}{(1+ uv)^{2}},
\end{equation}

we aim to get at least a $C^{0}$ metric, therefore there should be an appropriate matching condition along the shock wave between the two black holes. For instance, from the relationship between the ADM mass, Schwarzchild radius and the AdS scale, one has $R_{1}^{2} M = (M+E_{+})R_{2}^{2}$, where $R_{1}$ is the horizon radius as seen by the first asymptotic universe (where we send the shock wave from), and $R_{2}$ is the horizon radius for the second universe. Considering additional matching conditions\footnote{This includes for instance, the size of the circle that is at each point in the Penrose diagram, figure \ref{fig.}. This radius has to be continuous as we cross the shockwave.} for the metric of the two spacetimes will lead to the simple result, that past horizon of black hole 1 and future horizon of black hole 2 will miss each other (fig. \ref{fig.}) by an amount $ \frac{E_{+}}{4M} e^{\frac{2\pi t_{0}}{\beta}}$,
\begin{equation}
    v_{1} = v_{2} + \alpha, \; \text{ where} \; \alpha = \frac{E_{+}}{4M} e^{\frac{2\pi t_{0}}{\beta}}
\end{equation}
in the limit $\frac{E_{+}}{M} \rightarrow 0$, while keeping $\alpha $ fixed, therefore $t_{0} \rightarrow \infty$. The full metric will then be 

\begin{equation}
    ds^{2}= \frac{-2 \beta R_{2}/\pi \; dudv_{2} + R_{2}^{2}(1-u(v_{2}+\alpha \theta(u)
    ))^{2} d\phi^{2}}{(1+ u(v_{2}+\alpha \theta(u)
    ))^{2}},
\end{equation}

which reduces to the BTZ metric with ADM mass $M$ in the left universe and mass $M+E_{+}$ in the right universe. One important aspect of this back-reaction is visible in the computation of the geodesic distance, $d_{\text{trav}}$, between the two boundary points at $t_{1}$ and $t_{2}$, which we discussed in the previous section.   

\begin{equation}
    d_{\text{trav}}/l = 2\text{ log}\frac{2r}{R} + 2 \text{ log}\left( \text{ cosh}(\frac{\pi (t_{1}+t_{2})}{ \beta}) + \frac{\alpha}{2} e^{-\frac{\pi}{\beta} (-t_{1}+t_{2})} \right)
\end{equation}

where $r$ is the radial distance for each of the points at the two boundaries. It is easy to see that if one takes $t_{1}=t_{2}=0$, there is an $\alpha$ dependent contribution to the distance, making it longer (or smaller depending on the sign of $\alpha$). The effect is to decrease the correlation between the two CFTs, since the 1-2 correlation function is given by the exponential of minus the renomalized distance. In \cite{Shenker:2013pqa}, mutual information was also computed which is shown to be decreasing when the back reaction is significant.

One should also notice that, if one takes $t_{1} \sim -t_{0}$ while $t_{2}\sim t_{0}$, the back reaction is $O(G_{N})$ and vanishes in the limit $\frac{E_{+}}{M}$ goes to zero. This is to be expected since the big back-reaction in the wormhole geometry is due to the blue shift of the small perturbation sent into the black hole from the far past, and not because the energy of the initial perturbation itself was big. Therefore, for boundary measurements done not too far in the past or future away from the initial perturbation, there will be vanishing back-reaction in the semiclassical limit. On the other hand when there is a significant time difference from the shock wave (on the order of the scrambling time, $T \sim \frac{\beta}{2\pi} \text{ log}S$, $S$ being entropy of the black hole), which translates into significant boost difference in the bulk, it will lead to a big back reaction in the local frame of the measurement.  

The 1-2 boundary two point functions in the shock wave geometry, where the operators are located at $t_{1} = t_{2} = \phi_{1} = \phi = 0$ is given by, as discussed in the previous section,
\begin{equation}\label{shock 2 point}
  \langle O_{1}O_{2}\rangle_{SH} = \left( \frac{1}{1 + \frac{\alpha}{2}}\right)^{2ml}
\end{equation}

where $m$ is the mass of the bulk dual operator and $l= \sqrt{\frac{\beta R_{2}}{2\pi}}$.

However, the most relevant point for us at this point is that from the Einstein equations for the above back-reacted metric we have,

\begin{equation}
    \int_{-\infty}^{\infty} du \;T_{uu} = \frac{\alpha}{4\pi G_{N}}
\end{equation}

At this point it is important to note that sending a small perturbation of energy $E_{+}$ from the far past does not produce a traversable wormhole. In fact it has the opposite effect, making the the correlation between the two sides weaker. This can be seen above since for positive energy $E_{+}$, $\alpha$ is positive. However if one had a negative energy shock wave, it would violate ANEC and it would be possible to get a traversable wormhole. In this case, $\alpha$ becomes negative and the correlation between the two CFTs will get stronger until it diverges, signaling a causal contact \cite{Kawamoto:2025oko} between the two boundaries. It was emphasized in \cite{Maldacena:2017axo} that the creation of a small perturbation in the far past and measuring it in the far future (i.e, double trace deformation) has the effect of creating a negative energy shock wave. 

The basic feature that allowed for this effect is the time asymmetric nature of the evolution. In particular, if we considered the fluctuations on the left and right boundaries as emerging from the white hole region and evolve with a time independent Hamiltonian, the effect of the negative energy shock wave will be canceled by the secondary shock wave emerging from the white hole\footnote{This can be seen by considering a theory with just a scalar field and a state created by acting with two spacelike separated operators on the vacuum. The expectation value of the stress energy tensor is given to linear order in some coupling, that measures the strength of the fluctuation, by the commutator of the stress energy operator with the product the two operators. The result is localized along light cones where it is negative along one of the forward light cones and positive on the other and vice versa for the backward light cones. The ANEC operator, $\int_{-\infty}^{\infty} du \;T_{uu}$ along some in-extendable null path gets contribution from the positive forward light cone and negative backward light cone or vice versa, which cancel each other. Considering time dependent Hamiltonian essentially removes the backward light cones therefore it is possible to get a negative value for the expectation of the ANEC operator.}. As was emphasized in the GJW protocol however, the violation of ANEC was achieved by implementing the left and right fluctuations by a time dependent Hamiltonian where the system evolves with the usual Hamiltonian of the thermofield double state until a certain time $t_{0}$ ($-t_{0}$ for the left boundary) is reached, after which the system is evolved with a modified Hamiltonian, one with a double trace deformation.    

They primarily considered nearly AdS$_{2}$ geometries, that is extending the direct computation by Gao, Jafferis and Wall we reviewed earlier, and the main result, that is \eqref{C}, can be generalized to higher dimensional black holes. In particular, it was noted that the perturbation to the Hamiltonian that was considered in \eqref{pert. hamil}, can be taken to be an implementation shock wave scenario we discussed above but with a negative energy, as long as $t_{0}$ goes to infinity appropriately, i.e, keeping $\alpha$ constant. Let $gV(t)$ be the double trace deformation (like \eqref{pert. hamil}), where $g$ controls the strength of the interaction\footnote{The form of $gV(t)$ will be discussed more generally in the following section.}, then while we take $t_{0}$ to infinity\footnote{Note that $V(t) \neq 0 $ for $t\geq t_{0}$.}, we consider a probe particle that we send at $O(1)$ time from the boundary, far way from the shock wave. As opposed to the previous discussion, here the probe does not originate in the far past and emerge in the far future however it will have a time advance (and a time delay in the case of \eqref{shock 2 point} with positive energy) for the right sign of $g$ and will traverse the wormhole reaching the other boundary. 

To check this one can compute the response of an operator in CFT 1, from a unitary fluctuation of the CFT 2 creating the probe particle, in the far past from the time the coupling $V$ is turned on, in the presence of $e^{igV}$. A priori from acting with $gV(t)$, a unitary fluctuation is expected to not produce any effect on the second CFT since the two CFTs are causally disconnected. This is seen in the bulk by the fact that the probe particle will eventually end up in the black hole interior and the singularity\footnote{A non unitary operator however can produce some non trivial effect on the the other CFT as a result of the non zero entanglement between the two systems. This property is captured by the Reeh Schlider theorem for the thermofield double state.}. Therefore we compute the difference between the expectation value of an operator on CFT 1, in the presence of the unitary operator of CFT 2 and in the absence of the operator; while time dependent interaction is present, 

\begin{equation}
    \langle e^{-i\epsilon_{2}O_{2}} e^{-igV} O_{1}\; e^{igV}  e^{i\epsilon_{2}O_{2}}\rangle -  \langle e^{-igV} O_{1}\; e^{igV}\rangle \simeq -i\epsilon_{2} \langle [O_{2},O_{1}]\rangle_{V} + O(\epsilon_{2}^{2})
\end{equation}

for small $\epsilon_{2}$, where 

\begin{equation}\label{comm}
    \langle [O_{2},O_{1}]\rangle_{V} = 2\;\text{Im }C 
\end{equation}

and,

\begin{equation}
    C := \langle  e^{-igV} O_{1}(t_{1},\bar{x})\; e^{igV}  O_{2}(t_{2},\bar{x})\rangle.
\end{equation}

In writing \eqref{comm} we assumed $O_{1,2}$ are Hermitian single trace operators. In \cite{Maldacena:2017axo}, $C$ was computed to be 

\begin{equation}\label{C}
    C = \langle  e^{-igV} \rangle \int\;dp_{+} \langle O_{1}|p_{+}\rangle \langle p_{+}|O_{2}\rangle \text{ exp} \left[ ig \langle O^{'}_{2} e^{iG_{N}e^{\frac{2\pi t}{\beta}}p_{+} Q_{-}}O^{'}_{1}\rangle \rangle\right]  
\end{equation}

where $p_{+}$ is the null momentum for the $O_{1,2}$ while $Q_{-}$ is acting on the $O^{'}_{1,2}$ operators (the operators performing the double trace deformation) with basis $\ket{q_{-}}$. In particular, $C$ is found to be non zero, which implies the effect of a unitary operation on CFT 2 is recorded on CFT 1. Since physical operations an experimenter would do are implemented by a unitary operation, this implies a causal connection between the two CFT's.

The essential input in getting this result is the explicit scattering computations of terms that appear in the correlator, which is equivalent to the eikonal resummation of gravitational exchanges\cite{Kabat:1992tb,Verlinde:1991iu,tHooft:1987vrq,Amati:1987uf}. That is, the correlator like $\langle O_{1}O_{1}^{'}O_{2}^{'}O_{2}\rangle$ with the particular time ordering discussed above can be approximated by the integral over the momenta of the momentum wave functions for the particle created by each of the single trace operators multiplied by the phase factor, 
\begin{equation}
   e^{iG_{N}p_{+}q_{-}e^{\frac{2\pi t}{\beta}}}.
\end{equation}

Then \eqref{C} follows from the re-summation of such correlators in the $g$ expansion of $C$. As mentioned before, this formula is also valid for higher dimensional traversable wormholes \cite{Maldacena:2017axo,Ahn:2020csv} however for the general case, one has to do the replacement, $\ket{p_{+}} \rightarrow \ket{p_{+}, \bold{r}}$ and $\ket{q_{-}} \rightarrow \ket{q_{-}, \bold{r^{'}}}$. In addition, $G_{N} \rightarrow G_{N}f(\bold{r}-\bold{r^{'}})$, to include the transverse profile of the shock wave, where $\bold{r}$ denotes the traverse direction in higher dimensions.    
\\

\emph{The main goal of this paper is to recover \eqref{C} from a purely algebraic perspective without having to use the input from the gravitational scattering computations.} 
\\

As a concluding remark for this section, we would like to note that, starting from \eqref{C}, equation \eqref{shock 2 point} can be extended for the case of nearly AdS$_{2}$ geometries \cite{Maldacena:2017axo} by noticing that $Q_{-}$ is one of the $SL(2)$ symmetry generators and we have 

\begin{equation}
    \langle O^{'}_{2} e^{-iaQ_{-}}O^{'}_{1}\rangle = \left(\frac{1}{1+\frac{a}{2}}\right)^{2\Delta}
\end{equation}

thus the exponential inside the integral in \eqref{C} can be expanded in the small back reaction limit, that is taking $G_{N}e^{\frac{2\pi t}{\beta}}$ to be small while $gG_{N}e^{\frac{2\pi t}{\beta}}$ is constant. The correlator $C$ can then be re-written as follows (see section 2.3 of \cite{Maldacena:2017axo} for the more detailed steps), 

\begin{equation}
    C = \langle O_{1} e^{-ia^{+}P_{+}}O_{2}\rangle = \left(\frac{1}{1+\frac{a^{+}}{2}}\right)^{2\Delta}
\end{equation}

where $a^{+}= \frac{-\Delta \;g G_{N}e^{\frac{2\pi t}{\beta}}}{2^{2\Delta + 1}}$ and where $P_{+}$ is the second $SL(2)$ symmetry generator and the $p_{+}$'s are its eigenvalues.

\section{Algebraic traversable wormholes}\label{alg tra worm}

We consider the canonical ensemble, the thermofield double state described in the section \ref{sec:trave}. We use single trace operators in CFT$_{1}$ with no explicit factors of $N$, to construct a set of operators $\mathcal{M}$ by taking  bounded function of their $O(1)$ sum of $O(1)$ products. We have not taken the large $N$ limit and in particular, we take $N$ to be a very large but finite number. If an operator in our set is constructed out of a product of $k$ single trace operators, then $k$ is much less than $N^{2}$ for large enough $N$ since it's $O(1)$. As would be expected the Hamiltonian will not be an element of this set since it includes explicit factor of $N^{2}$. However the difference of the first CFT and the second CFT Hamiltonians generates a well defined unitary action on our set, in the sense that $e^{i(H_{1}-H_{2})}$ will map operators in our set to another operator in the set\footnote{An easy way to see this is by noting that the commutator of $H_{1}-H_{2}$ with any of the operator in our set is proportional to the derivative operator with respect to time acting on the same operator and thus also an element of the set.}. We can construct the would be modular operator using these Hamiltonians as\footnote{This is indeed the modular operator for the algebra of operators in the full system.},

\begin{equation}\label{mod}
    \Delta^{it} = e^{it\beta(H_{1}-H_{2})}
\end{equation}

From now on we will take $\beta$ to be one. We can also consider the modular conjugation operator $J$ of the full system. This will map the set $\mathcal{M}$ to its `commutant', i.e, the corresponding, similarly constructed set of operators in the second system which we call $\mathcal{M^{'}} =  J\mathcal{M}J$.

In addition, we define a subset $\mathcal{N}$ of $\mathcal{M}$, which includes operators similarly constructed but with $t>t_{0}$, for some $t_{0}>0$. Another way to describe these sets, for instance $\mathcal{M}$ is that it is exactly the subset of operators in CFT$_{1}$ whose large $N$ limit exists. Similarly, $\mathcal{N}$ is the exact subset operators in $\mathcal{M}$ whose large $N$ limit defines the subalgebra of operators in the large $N$ limit of the boundary theory, with $t>t_{0}$. The reason we are interested in these sets rather than their large $N$ algebras themselves is that, as we will see, the objects of our interest are better understood first at finite $N$, before taking the large $N$ limit. Traversable wormholes are distinct bulk configurations that are not accessible from the commonly taken large $N$ limit of thermofield double state. We will see how a slightly modified large $N$ limit of thermofield double state will give rise to traversable wormholes as its bulk configuration, starting with these sets. The reason these sets are not algebras yet is that if we take the product of enough of the operators in $\mathcal{M}$, then our defining condition will not be satisfied for the resulting operator (even though it will obviously be satisfied if we first take $N$ to infinity). However as subset of the full CFT$_{1}$, the modular conjugation operator will map the operators to operators in CFT$_{2}$, which commute with $\mathcal{M}$.

Even though these operators are related to the operators with $t<t_{0}$ by time evolution, since the Hamiltonian is not part of $\mathcal{M}$, the set $\mathcal{N}$ is a proper subset of $\mathcal{M}$. However, we have to be careful here since the renormalized Hamiltonian is an element of $\mathcal{M}$. But we can see that even if we use the renormalized Hamiltonian to evolve the operators, it will not produce $O(1)$ time translation, as long as we consider times of order one, i.e, $t<< N$. Therefore as soon as we consider $t\sim O(N)$, $\mathcal{N}$ will stop becoming a proper sub set and becomes the full set $\mathcal{M}$. For instance, if $a(2t_{0},\bar{x}) \in \mathcal{N}$, it can be written in terms of $a(\frac{1}{2} t_{0},\bar{x}) \notin \mathcal{N}$ by evolving it by $t= 3Nt_{0}/2$ with $e^{it\frac{\tilde{H_{1}}}{N}}$, $\tilde{H_{1}}$ is the subtracted Hamiltonian \cite{Witten:2021unn}. Therefore we impose that we will not evolve the operators in $\mathcal{M}$ by times of $O(N)$. This condition is reasonable for our purposes since eventually we are interested in taking the large $N$ limit where this set becomes a von Neumann algebra, and in this limit no finite evolutions can produce such time translations. 

On a slightly different note, we have not yet set the operators in $\mathcal{N}$ to be operators with $t \sim O(1)$ only. In other words, above we said the sub set $\mathcal{N}$ involves operators in $\mathcal{M}$ that are localized at time $t>t_{0}$, but we have not imposed the upper bound for $t$. In particular a problem arises when $t\sim Ne^{e^{S}}$, where $S$ is the entropy of CFT$_{1}$. Since $e^{e^{S}}$ time is the Poincare recurrence time and the CFT's have discrete energy eigenstates\footnote{Recall that we are taking the CFT's to be on $\mathbf{R} \cross \mathbf{S}^{d-1}$}, an operator $a(t\sim Ne^{e^{S}},\bar{x})$ is arbitrary close to $a(0,\bar{x})$ since,

\begin{equation}
    a(t,\bar{x}) = \sum_{ij} e^{i\frac{t}{N}(E_{i}-E_{j})} \bra{i}a(0,\bar{x})\ket{j} \;\ket{i}\bra{j}
\end{equation}

is quasiperiodic for discrete $E_{i,j}$. Therefore one defines $\mathcal{N}$ to be not just the operators in $\mathcal{M}$ with $t >t_{0}$, but also with an upper bound on $t$ given by $T \sim Ne^{e^{S}}$. Again no finite $t$ in the large $N$ limit will have access to such long times thus the upper bound will only have a finite $N$ effect.
%An alternative way to address this issue would be to just remove $\frac{\tilde{H}}{N}$ from our set $\mathcal{M}$.

Once we define the set $\mathcal{N}$ as above, we consider another set in the `commutant' $\mathcal{M}^{'}$, $J\mathcal{N}J$. Notice that $J\mathcal{N}J$ is not same as the set of all operators that commute with $\mathcal{N}$ since $J$ is not the modular conjugation operator for $\mathcal{N}$\footnote{Even in the large $N$ limit, $J$ will be the modular conjugation operator for $\mathcal{M}$ and not $\mathcal{N}$.}. The set $J\mathcal{N}J$ is in fact the subset of operators in $\mathcal{M}^{'}$ with $t<-t_{0}$, since $J$ is anti unitary. 

Now we take the set,
$
    \mathcal{N}\cup J\mathcal{N}J
$
which is the set constructed from the single trace operators in $ \mathcal{N}$ and $J \mathcal{N}J$ and taking the bounded functions of their arbitrary $O(1)$ products and sum. Using the modular operator we can construct a chain of nested sets. For instance the set\footnote{A more mathematically precise discussion on similarly nested algebras can be found in appendix \ref{A}}
\begin{equation}
    \sigma_{t_{1}}(\mathcal{N}\cup J\mathcal{N}J) := \Delta^{-it_{1}}(\mathcal{N}\cup J\mathcal{N}J)\Delta^{it_{1}}
\end{equation}

for $t_{1}>0$ is a subset of $\mathcal{N}\cup J\mathcal{N}J$ which are constructed out of operators in $\mathcal{M}$ with $t>t_{0}+t_{1}$ and operators in $\mathcal{M}^{'}$ with $t<-t_{0}-t_{1}$. More generally, if $t_{1}<t_{2}$, then we have,
$   \sigma_{t_{2}}(\mathcal{N}\cup J\mathcal{N}J) \subset \sigma_{t_{1}}(\mathcal{N}\cup J\mathcal{N}J)$.

\begin{figure}

\begin{tikzpicture}[scale=3.8]
  \message{Extended Penrose diagram: Schwarzschild black hole^^J}
  
  \def\R{0.08} % size lightcone
  \def\Nlines{3} % number of world lines (at constant r/t)
  \pgfmathsetmacro\ta{1/sin(90*1/(\Nlines+1))} % constant r/t value 1
  \pgfmathsetmacro\tb{sin(90*2/(\Nlines+1))}   % constant r/t value 2
  \pgfmathsetmacro\tc{1/sin(90*2/(\Nlines+1))} % constant r/t value 3
  \pgfmathsetmacro\td{sin(90*1/(\Nlines+1))}   % constant r/t value 4
  \coordinate (-O) at (-1, 0); % center III: origin (r,t) = (0,0)
  \coordinate (-S) at (-1,-1); % south III: t=-infty, i-
  \coordinate (-N) at (-1, 1); % north III: t=+infty, i+
  \coordinate (-W) at (-1, 0); % east III:  r=-infty, i0
  \coordinate (L) at (0.55, 0.55); % N, 
  \coordinate (M) at (1, 0.1); % N, 
  \coordinate (-E) at ( 0, 0); % west III:  r=+infty, i0
  \coordinate (O)  at ( 1, 0); % center I: origin (r,t) = (0,0)
  \coordinate (S)  at ( 1,-1); % south I: t=-infty, i-
  \coordinate (N)  at ( 1, 1); % north I: t=+infty, i+
  \coordinate (-L) at (-0.55, -0.55); % JNJ,
   \coordinate (-M) at (-1, -0.1); % N,
  \coordinate (E)  at ( 1, 0); % east I:  r=-infty, i0
  \coordinate (W)  at ( 0, 0); % west I:  r=+infty, i0
  \coordinate (B)  at ( 0,-1); % singularity bottom
  \coordinate (T)  at ( 0, 1); % singularity top
  \coordinate (X0) at ({asin(sqrt((\ta^2-1)/(\ta^2-\tb^2)))/90},
                       {-acos(\ta*sqrt((1-\tb^2)/(\ta^2-\tb^2)))/90}); % particle 1
  \coordinate (X1) at ({asin(sqrt((\tc^2-1)/(\tc^2-\td^2)))/90},
                       {acos(\tc*sqrt((1-\td^2)/(\tc^2-\td^2)))/90}); % particle 2
  \coordinate (X2) at (45:0.87); % particle falling in BH horizon
  \coordinate (X3) at (0.60,1.05); % particle falling in BH singularity
  
  \begin{scope}
    
    % CLIP to fill inside zigzag lines
    \clip[decorate,decoration={zigzag,amplitude=2,segment length=6.17}]
      (S) -- (-S) --++ (-1.1,-0.1) |-++ (4.2,2.2) |- cycle;
    \clip[decorate,decoration={zigzag,amplitude=2,segment length=6.17}]
      (-N) -- (N) --++ (1.1,0.1) |-++ (-4.2,-2.2) |- cycle;
    
    % REGIONS FILLS
    \fill[mylightpurple] (-N) |-++ (2,0.1) -- (N) -- (-S) -- (S) -- cycle;
    \fill[mylightpurple] (-S) |-++ (2,-0.1) -- (S) -- (-N) -- (N) -- cycle;
    
    \fill[mylightblue] (-N) -- (-E) -- (-S) -- (-W) -- cycle;
    \fill[mylightblue] (N) -- (E) -- (S) -- (W) -- cycle;
     \fill[mygreen] (N) -- (L) -- (M) -- cycle;
      \fill[mygreen] (-S) -- (-L) -- (-M) -- cycle;
    
    % CONE BACK
   % \coneback{X0};
    %\coneback{X1};
    %\coneback{X2};
    
    % WORLD LINES
    \draw[world line] (-N) -- (-S) (N) -- (S);
    \draw[world line t] (-W) -- (-E) (W) -- (E) (0,-1.1) -- (0,1.1);
    \message{Making world lines...^^J}
    \foreach \i [evaluate={\c=\i/(\Nlines+1); \cs=sin(90*\c);}] in {1,...,\Nlines}{
      \message{  Running i/N=\i/\Nlines, c=\c, cs=\cs...^^J}
      %\draw[world line t,samples=2*\Nsamples,smooth,variable=\x,domain=-2:2] % region I/III, constant t
       % plot(\x,{-kruskal(\x*pi/4,\cs)})
       % plot(\x,{ kruskal(\x*pi/4,\cs)});
      \draw[world line,samples=\Nsamples,smooth,variable=\y,domain=0:2] % region I/III, constant r
       
        plot({-1+kruskal(\y*pi/4,\cs)},\y-1)
        plot({1-kruskal(\y*pi/4,\cs)},\y-1);
        %plot({1+kruskal(\y*pi/4,\cs)},\y-1)
         %plot({-1-kruskal(\y*pi/4,\cs)},\y-1)
     % \draw[world line,samples=\Nsamples,smooth,variable=\x,domain=0:2] % region II/IV, constant r
       % plot(\x-1,{kruskal(\x*pi/4,\cs)-1})
      %  plot(\x-1,{1-kruskal(\x*pi/4,\cs)});
     % \draw[world line t,samples=\Nsamples,smooth,variable=\y,domain=-1.05:1.05] % region II/IV constant t
      % plot({-kruskal(\y*pi/4,\cs)},\y)
       % plot({ kruskal(\y*pi/4,\cs)},\y);
        
    }

    % PARTICLE WORLD LINE
   % \draw[particle,decoration={markings,mark=at position 0.16 with {\arrow{latex}},
    %                                   mark=at position 0.45 with {\arrow{latex}},
     %                                   mark=at position 0.72 with {\arrow{latex}},
      %                                  mark=at position 0.90 with {\arrow{latex}}},postaction={decorate}]
      %(S) to[out=77,in=-70] (X0) to[out=110,in=-80] (X1)
       %   to[out=100,in=-90] (X2) to[out=75,in=-80] (X3);
    
  \end{scope}
  
  % BOUNDARIES
  \draw[singularity] (-N) -- node[above] {} 
  (N);
  \draw[singularity] (S) -- node[below] {} (-S);
  \path (S) -- (W) node[mydarkblue,pos=0.50,below=-2.5,rotate=-45,scale=0.85]
 {\contour{mylightpurple}{}};
  \path (L) -- (M) node[mydarkblue,pos=0.40,below=-2.5,rotate=-45,scale=0.85]
   {\contour{mylightpurple}{}};
  \path (W) -- (N) node[mydarkblue,pos=0.32,above=-2.5,rotate=45,scale=0.85]
    {\contour{mylightpurple}{}};
  \draw[thick,mydarkblue] (-N) -- (-E) -- (-S) -- (-W) -- cycle;
  \draw[thick,mydarkblue] (N) -- (L) -- (M) -- cycle;
  \draw[thick,mydarkblue] (-S) -- (-L) -- (-M) -- cycle;
  \draw[thick,mydarkblue] (N) -- (E) -- (S) -- (W) -- cycle;
  
  % REGIONS
  %\node[fill=mylightblue,inner sep=2] at (-O) {};
  %\node[fill=mylightblue,inner sep=2] at (O) {};
  %\node[fill=mylightpurple,inner sep=2] at (0,0.64) {};
  %\node[fill=mylightpurple,inner sep=2] at (0,-0.64) {};
  
  % INFINITY LABELS
  \node[above=1,left=1,mydarkblue] at (-1.2,0.5) {$\mathcal{M^{'}}$};
  \node[above=1,right=1,mydarkblue] at (1,0) {$t=0$};
  \node[above=1,right=1,mydarkblue] at (1.2,-0.5) {$\mathcal{M}$};
   \node[above=0.5,right=1,mydarkblue] at (1,0.5) {$\mathcal{N}$};
    \node[left=1,below=0.5,mydarkblue] at (-1.2,-0.5) {$J\mathcal{N}J$};
    \node[left=1,below=0.5,mydarkpurple] at (-1.2,-1) {CFT$-1$};
    \node[above=1,right=1,mydarkpurple] at (1,-1.05) {CFT$-2$};
  %\node[right=1,below=1,mydarkpurple] at (-S) {$i^-$};
  %\node[right=1,above=1,mydarkpurple] at (-N) {$i^+$};
  %\node[right=1,below=1,mydarkpurple] at (S) {$i^-$};
  %\node[right=1,above=1,mydarkpurple] at (N) {$i^+$};
  %\node[mydarkblue,below left=-1] at (-1.5,-0.5) {$\calI^-$};
  %\node[mydarkblue,above left=-1] at (-1.5,0.5) {$\calI^+$};
  %\node[mydarkblue,above right=-1] at (1.5,0.5) {$\calI^+$};
  %\node[mydarkblue,below right=-1] at (1.5,-0.5) {$\calI^-$};
  
  % LIGHT CONE FRONT
  %\conefront{X0};
  %\conefront{X1};
  %\conefront{X2};
  
  % ESCAPING PHOTONS
  %\draw[photon] (X0) ++ (45:0.1) --++ (45:0.3);
  %\draw[photon] (X1) ++ (45:0.1) --++ (45:0.3);
  
\end{tikzpicture}
\caption{$\mathcal{N}$ is a subalgebra of $\mathcal{M}$ including operators with $t>t_{0}$ for some positive $t_{0}$, while $J\mathcal{N}J$ is a subalgebra of $\mathcal{M}^{'}$ including operators with $t<-t_{0}$. Under modular evolution by a positive parameter $s$, $\mathcal{N}$ is translated into an even smaller sub algebra of $\mathcal{N}$ and $\mathcal{M}$ ans similarly for $J\mathcal{N}J$. $J\mathcal{N}J$ is translated into a smaller subalgebra because modular evolution by a positive parameter translates operators into the past for CFT-1.}\label{fig.hsmi}
\end{figure}
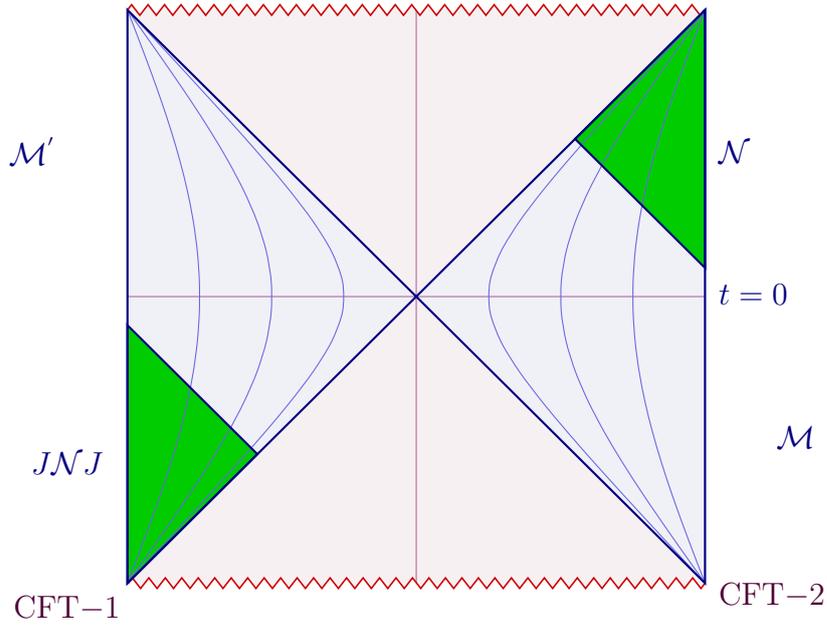

In particular, we can consider 
\begin{equation}
  \mathcal{A}_{i}= \cap_{t_{i}<t<0} \;\sigma_{t}(\mathcal{N}\cup J\mathcal{N}J). 
\end{equation}
Because of the nested structure just mentioned, we have 
\begin{equation}
 \cap_{t_{i}\leq t<0} \;\sigma_{t}(\mathcal{N}\cup J\mathcal{N}J) = \sigma_{t_{i}}(\mathcal{N}\cup J\mathcal{N}J)
\end{equation}
We have not put a bound on $t_{i}$, which can be $O(1)$, order the scrambling time or on the order of the Poincare recurrence time. However for our purposes, we shall proceed as follows. Let,
\begin{equation}
  \mathcal{A}_{N}= \cap_{t_{N}<t<0} \;\sigma_{t}(\mathcal{N}\cup J\mathcal{N}J) 
\end{equation}
where $t_{N}$ goes to infinity like $\frac{1}{2\pi}\text{log} (\frac{\gamma M}{p_{+}})$. where $p_{+}$ is the momentum of an initial infalling perturbation in its local frame, and $p_{+}\sim E_{+}$ if it is sent from the boundary in the at very early times (check section \ref{shock trav}). In particular, we want to to take $\frac{p_{+}}{M}$ to zero as $t_{N}$ goes to infinity, so that $\gamma =\frac{E_{+}}{M}e^{2\pi t_{N}} \sim E_{+}G_{N}e^{2\pi t_{N}}$ is kept constant. We aim to compute expectation values of operators in $\mathcal{A}_{N}$ in the large $N$ limit for states of order one excitation on top of the eternal black hole 
%and we will not take into account the back reaction of these excitations
. 
%In the case where one keeps track of the back reaction, $t_{N}$ will be shifted by $\frac{1}{2\pi}\text{ log}(p_{+})$, where $p_{+}$ is the momentum of the initial infalling momentum in its local frame. 
The set of operators $\mathcal{A}_{N}$ is still non-trivial since the upper bound of $t$ with respect to which $\mathcal{N}$ is defined, is the Poincare recurrence time; which is parametrically larger than $t_{N}$ for large $N$. For instance, if $\varphi_{1}(x_{0},\bar{x})$ and $\varphi_{2}(-y_{0},\bar{y})$ are single trace operators in CFT$_{1}$ and CFT$_{2}$, with $x_{0},y_{0} > t_{0}$; an element of $\mathcal{A}_{N}$ would be a bounded function of
\begin{equation}
    L_{N} = \int dxdy\; f(x)g(y) \varphi_{1}(x_{0}+t_{N},\bar{x})\varphi_{2}(-y_{0}-t_{N},\bar{y})
\end{equation}
where $f$ and $g$ are the appropriate smearing functions. For a more general function $h$ depending on $x$ and $y$, one can also write $L_{N} = \sigma_{t_{N}}(L_{0})$, where 
\begin{equation}\label{double trace}
    L_{0} = \int dxdy\; h(x,y) \varphi_{1}(x_{0},\bar{x})\varphi_{2}(-y_{0},\bar{y})
\end{equation}
Ultimately, what we are interested in is the large N limit. In this limit, the double commutant\footnote{The double commutant is taken within $\mathcal{B(H_{\text{trav}})}$, check section \ref{hilbert}.} of $\mathcal{M}$ will be a von Neumann algebra so does the double commutant of $\mathcal{N}$ \cite{Leutheusser:2022bgi,Leutheusser:2021frk} (with some abuse of notation we will again call these algebras $\mathcal{M}$ and $\mathcal{N}$ respectively). In fact, in the extreme large $N$ limit or $\frac{p_{+}}{M} \rightarrow 0$ limit, we expect these two to form half sided modular inclusions. The modular operator introduced above \eqref{mod} will be the modular operator for the algebra $\mathcal{M}$ and modular transformations for positive modular parameter will not take operators out of $\mathcal{N}\subset \mathcal{M}$. In addition, since $\mathcal{M}$ is the usual algebra of single trace operators discussed in \cite{Leutheusser:2022bgi,Leutheusser:2021frk}, it is clear that the `large $N$ limit of $\ket{TFD}$' will be a cyclic and separating state for it. For the algebra $\mathcal{N}$, however, a rigorous proof that the state is cyclic and separating is absent at the moment. We rather take this condition as an assumption, similar to the assumption taken for the semi-infinite time band sub-algebra considered by \cite{Leutheusser:2022bgi}, as they construct a timelike evolution operator into the interior of the eternal black hole. Therefore we have a state which is cyclic and separating for both $\mathcal{N}$ and $\mathcal{M}$, which act on a common GNS Hilbert space $\mathcal{H}$ and (see fig. \ref{fig.hsmi})
\begin{equation}
    \Delta^{-is}\mathcal{N}\Delta^{is} \subset \mathcal{N} \text{, for } s\geq0.
\end{equation}

Therefore according to Brochers et. al. \cite{Borchers:1991xk,Wiesbrock:1992mg,Borchers:2000pv}, there exists a one parameter unitary group $U(a)$, $a \in \mathbf{R}$, with generator $P = \frac{1}{2\pi}(\text{log } \Delta_{\mathcal{N}} - \text{log } \Delta ) \geq 0$ \footnote{By this statement we mean that for any two vectors in the Hilbert space, the matrix element of $P$ is positive. This is so since $\mathcal{N} \subset \mathcal{M}$ and therefore, $\Delta_{\mathcal{N}} \geq \Delta$, similarly $f(\Delta_{\mathcal{N}})\geq f(\Delta)$ for a monotone function $f$, such as the log function.}, where $\Delta_{\mathcal{N}}$ is the modular operator of $\mathcal{N}$ and,

\begin{enumerate}\label{prop of U}
    \item $ \Delta^{-is} U(a) \Delta^{is} = U(e^{2\pi s}a)$
    \item $ \mathcal{N} = U(-1) \mathcal{M} U(1)$
    \item $ J U(a)J = U(-a) $.
\end{enumerate}

where $J$ is the modular conjugation associated with the algebra $\mathcal{M}$. This group $U(a)= e^{-iaP}$ is called (-)half sided modular translations (-hsmt) and we define $\alpha_{a}(\mathcal{N}) := U^{\dagger}(a)\mathcal{N}U(a)$. Here we only listed properties that will be useful for us, a more complete list of properties of $U(a)$ with their derivations can be found in the above references.

\subsection{The action of half sided modular translations}

 To gain some physical intuition into these objects and to see how they act in the eternal black hole geometry, let's discuss what they look like in Minkowski spacetime in $d$ dimensions. Consider dividing the spacetime into $\mathbf{R}^{1,1} \cross \mathbf{R}^{d-2}$ and take the metric,
 \begin{equation}
     ds^{2}= -dt^{2} + dx^{2} + d\bar{z}^{2}
 \end{equation}
where $(t,x)$ corresponds to the coordinates of $\mathbf{R}^{1,1}$ and $\bar{z}$ to the Euclidean space $\mathbf{R}^{d-2}$. Take the $t=0$ Cauchy slice where we prepare the vacuum $\ket{\Omega}$ of the quantum field theory. Then we call the left Rindler wedge the region with $x < -|t|$, and the right Rindler wedge, the region with $x > |t|$. These are the domains of dependence for the $x<0$ and $x>0$ subregions of the Cauchy slice. 

Focusing in on the $\mathbf{R}^{1,1}$ factor of the spacetime, we have the momentum vector components $(P^{0},P^{1})$, where $P^{0}$ is the Hamiltonian of the theory. In particular, in the light cone coordinates,
\begin{equation}
    x^{\pm} = t \pm x \; \text{, and }\; P^{\pm} = \frac{1}{2}(P^{0} \pm P^{1} ). 
\end{equation}
In these coordinates, the right Rindler wedge is $x^{+}>0,\; x^{-}<0$ and the left wedge is $x^{+}<0,\; x^{-}>0$. If we take $K$ to be the boost operator about the origin, i.e, the origin is the fixed point of the boost generator, then we have,
\begin{equation}\label{boost mom rel}
    [K,P^{\pm}] = \pm i P^{\pm} \; \text{, and }\; e^{iKs}P^{\pm}e^{-iKs}=e^{\mp s}P^{\pm}
\end{equation}
If we take the algebra of operators in the right Rindler wedge to be the von Neuamnn algebra $\mathcal{M}$ that we considered earlier, then it is easy to show that the boost operator $K$ is in fact proportional to $-$log$\Delta$. 

In the approximation that we can factorize the right and left Rindler wedges by a factorized Hilbert space $\mathcal{H}_{L}\otimes \mathcal{H}_{R}$, we can compute the modular operator by computing the left and right density matrices, $\rho_{R}, \rho_{L}$ for the vacuum, since 
\begin{equation}
    \Delta = \rho_{R}\otimes\rho^{-1}_{L}
\end{equation}
In the continuum limit where the above factorization of Hilbert spaces is not possible, the above combination is well defined while the individual density matrices are not. In any case, the right density matrix for instance can be computed using the Euclidean path integral from the field degrees of freedom on the $x>0$ slice to different sets of fields on the same subregion of the Cauchy slice, which is really just a $2\pi$ rotation in the $(t_{E},x)$ space. Transforming back to the Lorentzian spacetime, the rotation operator becomes the boost transformation operator that acts on the right wedge, with boost parameter $-2\pi i$. Similarly, for the left density matrix will be given by the unitary boost operator acting on the left wedge.
\begin{equation}
    \Delta = e^{-2\pi K_{R}}e^{2\pi K_{L}} = e^{-2\pi K}
\end{equation}
Since this combination is well defined in the continuum limit, the $\Delta$ in the continuum limit is also given as $e^{-2\pi K}$. A more precise derivation involves the holomorphic properties of $e^{-2\pi is K}$ acting on a general vector in the Hilbert space of the quantum field theory, which is just (or can be approximated arbitrarily well enough by) bounded function of the smeared field operators in the right wedge acting on the vacuum, by Reeh Schlieder theorem \cite{Reeh:1961ujh,Witten:2018zxz}. Since these vectors are holomorphic in the strip $1/2 > \text{Im }s >0$, and continuous as Im$s$ approaches $1/2$, we can set the boost parameter $s = i/2$, which can be seen to reproduce the action of $\Delta^{-1/2}$. Therefore we conclude that the modular operator in Rindler spacetime is the boost operator. This is important for us since the regions close to the horizon of the eternal black hole in AdS looks like Rindler wedges in AdS. In particular, very close to the horizon of the eternal black hole in AdS, the modular operator implements boost transformations.

We take the small algebra $\mathcal{N}$ to be the region $x^{+}> 1,\; x^{-}<0$, the corresponding modular operator is just given in terms of the boost generator that preserves the point $x^{+}= 1,\; x^{-}=0$, which is related by translation to the original boost transformation. That is,
\begin{align}
    e^{isK_{\mathcal{N}}} &= e^{-ia_{\mu}P^{\mu}} e^{isK} e^{ia_{\mu}P^{\mu}} \\
    &= e^{iP^{-}} e^{isK}e^{-iP^{-}}
\end{align}
where $a^{\mu} = (a^{+},a^{-})= (-a_{-},a_{+}) = (1,0)$. Therefore we have 
\begin{equation}
    K_{\mathcal{N}} = K - P^{-}
\end{equation}
But by the argument given in the previous paragraph we have $K = -\frac{1}{2\pi}\text{log} \Delta$. Similar arguments also imply that $K_{\mathcal{N}}= -\frac{1}{2\pi}\text{log} \Delta_{\mathcal{N}}$. 

The generator of our half sided modular translations $U(a)$, is given by 
\begin{equation}
    P = \frac{1}{2\pi}(\text{log} \Delta_{\mathcal{N}} - \text{log} \Delta) =  P^{-}
\end{equation}
which is just the null translation generator in the increasing $x^{+}$ direction. In addition from \eqref{boost mom rel}, we see that 
\begin{equation}
    e^{-is \text{log}\Delta} \;P \;e^{is \text{log}\Delta} = e^{2\pi s}P \; \text{, and } \mathcal{N}= e^{iP} \mathcal{M}e^{-iP}
\end{equation}
 which imply property $(1)$ and $(2)$ in the properties of half sided modular translations we listed above.

Again, a crucial point that will important for us is that the generator of $U$, $P = P^{-}$ is the null translation generator in the $x^{+}$ direction for Rindler spacetime. In addition, since the eternal black hole geometry looks like Rindler wedges close to the horizon, the generator of half side modular translations acts like the null translation generator very close to the horizon.

For the sake of completeness, we would like to mention also that there is another (+)half sided modular translation (+hsmt) for a different $\mathcal{N}_{-} \subset \mathcal{M}$, with $\Delta^{-is} \mathcal{N}_{-}\Delta^{is} \subset \mathcal{N}_{-}$ for $s\leq0$. In this case, we have $P = P^{+}$ which is the null translation in the $x^{-}$ direction. In addition, the $e^{2\pi s}$ factor in property $(1)$ of $U(a)$'s, will be replaced by $e^{-2\pi s}$; while $\mathcal{N}_{-} = U(1)\mathcal{M}U(-1)$. 

\subsection{Algebraic computation}

Coming back to the eternal black hole discussion, because of property (1) of -hsmt, we have $\mathcal{N}=\alpha_{1}(\mathcal{M})$. On the other hand, because of property (3), we have $J\mathcal{N}J=U(1)\mathcal{M}'U(-1)=\alpha_{-1}(\mathcal{M}')$. In the extreme large $N$ limit, the von Neumann algebra generated by the set $\mathcal{N}\cup J\mathcal{N}J$ is 

\begin{equation}
   (\mathcal{N} \cup J \mathcal{N}J)'' = \mathcal{N}\vee J\mathcal{N}J =  \alpha_{1}(\mathcal{M}) \vee \alpha_{-1}(\mathcal{M}')
\end{equation}
 
 As stated before, the kinds of operators present in $\mathcal{A}_{N}$ are $L_{N}=\sigma_{t_{N}(L_{0})}$, where $L_{0}$, \eqref{double trace}, is a certain smeared double trace operator in $\mathcal{N} \cup J \mathcal{N}J$. That is, we have

 \begin{equation}
     L_{N}=\sigma_{t_{N}}(L_{0})=\sigma_{\frac{1}{2\pi}\text{log}\gamma\; +\frac{1}{2\pi}\text{log}\frac{M}{p_{+}}}(L_{0})= \sigma_{\frac{1}{2\pi}\text{log}\frac{M}{p_{+}}}(\;\sigma_{\frac{1}{2\pi}\text{log}\gamma} (L_{0})\;)
 \end{equation}

 Notice that in the large $N$ limit, $\frac{M}{p_{+}} \sim \frac{1}{G_{N}}$, which goes like $N^{2}$ for AdS$_{5}$/CFT$_{4}$ for instance. In particular, log$\frac{M}{p_{+}}$ goes like log$(N^{2})$ while $\gamma$, and so log$\gamma$, is kept a fixed constant. Thus in the large $N$ limit ,
 \begin{equation}
     L_{N} = \sigma_{\frac{1}{2\pi}\text{log}N^{2}} (\;\sigma_{\frac{1}{2\pi}\text{log}\gamma} (L_{0})\;)
 \end{equation}
 In other words, the operator that we have in the extreme large $N$ limit is,
 \begin{equation}
     L = w-\lim_{s \rightarrow \infty} \sigma_{s} (\;\sigma_{\frac{1}{2\pi}\text{log}\gamma} (L_{0})\;)
 \end{equation}
 where $w-\lim$ is the weak limit, in the sense that the matrix elements of $\sigma_{s} (\;\sigma_{\frac{1}{2\pi}\text{log}\gamma} (L_{0})\;)$ above converge to $L$ as $s$ goes to infinity, and we define $s$ to go to infinity like $\frac{1}{2\pi}\text{log}N^{2}$.
\\

 The central claim is that \emph{acting with the bounded functions of $L_{N}$ on the thermofield double state as we take the large N limit, will make the eternal black hole a traversable wormhole, in the sense of Gao-Jafferis-Wall\cite{Gao:2016bin}.}
\\

 To this end we consider $N$ to be `extremely large' but finite (say $10^{10^{10}}$) and perturb our thermofield double state on CFT$_{1}$ by a unitary perturbation therefore with no other operators acting, the perturbation will not be detected from CFT$_{2}$. However, if we in addition act with a bounded function of $L_{N}$, then we will find that the expectation value of an operator in CFT$_{2}$ will depend on the unitary perturbation present in the CFT$_{1}$. Therefore, we consider the state

 \begin{equation}
     \ket{\Phi} = e^{igL_{N}} e^{i\epsilon_{1}B_{1}} \ket{TFD}
 \end{equation}
 
where $B_{1}$ is a bounded function of appropriately smeared single trace operator in $\mathcal{M}$. The time at which we apply with $B_{1}$ will not depend on $N$ while, for $L_{N}$, there will be a time dependence on the order of log$N^{2}$ as discussed above. If $B_{2}$ is a bounded function of single trace operators in $\mathcal{M}^{'}$, the relevant object we are computing is 

\begin{equation}
    \langle \Phi| B_{2} |\Phi \rangle -  \langle TFD| e^{-igL_{N}} B_{2} e^{igL_{N}} | TFD \rangle
\end{equation}
in the large $N$ limit.

This difference being dependent on $\epsilon_{1}$ implies that $ \langle \Phi| B_{2} |\Phi \rangle$ depends on the strength of the unitary perturbation on CFT$_{1}$. Following the same steps as in the previous section we see that, with 

\begin{equation}\label{alg C}
    C = \langle e^{-igL_{N}} B_{1}(t_{1},\bar{x}) e^{igL_{N}}B_{2}(t_{2},\bar{x})\rangle,
\end{equation}

the above difference is given by $2 \text{ Im}C$ up to $O(\epsilon_{1}^{2})$ corrections. Therefore the computation of \eqref{alg C} in the large N limit includes all the relevant information. We notice that,
\begin{align*}
    C =& \langle e^{-igL_{N}} B_{1}(t_{1},\bar{x}) e^{igL_{N}}B_{2}(t_{2},\bar{x})\rangle \\
    =& \langle \sigma_{t_{N}}(e^{-igL_{0}}) B_{1}(t_{1},\bar{x}) e^{igL_{N}}B_{2}(t_{2},\bar{x})\rangle \\
    =& \langle e^{-igL_{0}} \Delta^{it_{N}} B_{1}(t_{1},\bar{x}) e^{igL_{N}}B_{2}(t_{2},\bar{x})\rangle
\end{align*}
since $\Delta$ annihilates $\ket{TFD}$. But we have as we take $N$ to infinity, $t_{N}$ goes to infinity, and

\begin{equation}
   w- \lim_{t \rightarrow \pm\infty} \Delta^{it} = P_{\Psi}
\end{equation}
which follows from the uniqueness of the the thermofield double state. In particular, the generator log$\Delta$ has zero eigenvalue only for the thermofield double state, and it has continuous strictly positive or strictly negative eigenvalues on any other state. We are employing this argument here since the difference between successive energy eigenstates goes like $O(e^{-N^{2}})$ and times of order log$N^{2}$ will not be able to resolve the microstates. Then, the above result follows from Riemann–Lebesgue lemma \cite{longo,Lechner:2021fvy}.     

Therefore one proceeds as follows, \footnote{For the case of SYK, where we take a sum of a large number of double trace operators with finite coefficient, it can be showed that, \cite{Maldacena:2017axo}, $\langle e^{igL_{0}}\rangle = e^{ig\langle L_{0}\rangle}$ },

\begin{equation}
    C=\langle e^{igL_{0}}\rangle \langle B_{1}(t_{1},\bar{x}) e^{igL}B_{2}(t_{2},\bar{x})\rangle + O(1/N)
\end{equation}

where we assume that at large but finite $N$, the correlation function $C$ is given by the $N=\infty$ leading term and $O(1/N)$ corrections to this leading term with a possible additional $O(e^{-N^{2}})$ non perturbative correction, as is standard in computations of correlators in AdS/CFT. Note also that the above property of $\Delta^{it}$ is shown for type III$_{1}$ algebras which are emergent in the extreme large $N$ limit. Intuitively, in this order of operators, fluctuations separated by $t_{N}$ have no correlation since $t_{N}$ is much smaller that for instance the Poincare time.

The next step is to compute $$ \langle B_{1}(t_{1},\bar{x}) e^{igL}B_{2}(t_{2},\bar{x})\rangle= \lim_{N\rightarrow \infty} \langle B_{1}(t_{1},\bar{x}) e^{igL_{N}}B_{2}(t_{2},\bar{x})\rangle$$. For that we need to expand the exponential on the right hand side at the n$^{th}$ order, we consider 

\begin{equation}
   \Theta_{n} = (ig)^{n} \langle B_{1}(t_{1},\bar{x}) \left[ \int dxdy \;h(x,y) \sigma_{t_{N}}(\varphi_{1}(x_{0},\bar{x})\varphi_{2}(-y_{0},\bar{y}))\right ]^{n}B_{2}(t_{2},\bar{x})\rangle
\end{equation}
when $n=1$, we have 

\begin{align*}
    \Theta_{1} = ig \int dxdy \;h(x,y) \langle B_{1}(t_{1},\bar{x})   \sigma_{t_{N}}(\varphi_{1}(x_{0},\bar{x})\varphi_{2}(-y_{0},\bar{y}))B_{2}(t_{2},\bar{x})\rangle
\end{align*}

Note that $\varphi_{1,2} \in \mathcal{N},J\mathcal{N}J$ which in the large $N$ limit, is given by $\alpha_{1}(\mathcal{M}), \alpha_{-1}(\mathcal{M}^{'}) $.

\begin{align*}
    \Theta_{1} = ig \int dxdy \;h(x,y) \langle B_{1}(t_{1},\bar{x})   \sigma_{s +\frac{1}{2\pi}\text{log}\gamma} \left(\;\alpha_{1}(D^{0}_{1}(x_{0},\bar{x}))\alpha_{-1}(D^{0}_{2}(-y_{0},\bar{y}))\;\right)B_{2}(t_{2},\bar{x})\rangle + O(1/N),
\end{align*}

 as $N \rightarrow \infty$, where $D^{0}_{1,2}$ is a certain bounded function of smeared single trace operators in $\mathcal{M, M^{'}}$ respectively. Again, it is only in the large $N$ limit, $\mathcal{M,N,M^{'}}$ become fully fledged von Neumann algebras and therefore $\alpha(.)$ is defined. The first term above corresponds to the extreme large $N$ limit, which is the leading contribution for the \eqref{alg C} where $N$ is very large but finite. 

Forgetting the coordinates to simplify the notation we have, in the large $N$ limit,
\begin{align*}
     \Theta_{1} = & ig \int dxdy \;h \langle B_{1}  \sigma_{s +\frac{1}{2\pi}\text{log}\gamma} \left(\;\alpha_{1}(D^{0}_{1})\alpha_{-1}(D^{0}_{2})\;\right)B_{2})\rangle + O(1/N),\\
     = & ig \int dxdy \;h \langle B_{1}  \sigma_{s} \left(\;\alpha_{\gamma}(D_{1})\alpha_{-\gamma}(D_{2})\;\right)B_{2})\rangle + O(1/N),
\end{align*}
where in the last step we used the fact that $\sigma_{t}(\alpha_{s}(.)) = \alpha_{s\; e^{2\pi t}}(\sigma_{t}(.))$ following from property 1 of the properties of (-)hsmt discussed in the previous section. Therefore, $D_{1,2}= \sigma_{\gamma}(D^{0}_{1,2}) \in \mathcal{M, M^{'}}$.  

We then have,
\begin{align*}
     \Theta_{1} = & ig \int dxdy \;h \langle B_{1}\;  \Delta^{-is} U^{\dagger}(\gamma)\;D_{1}\;U(2\gamma) \;D_{2}\; U(-\gamma) \Delta^{is}\;B_{2})\rangle + O(1/N),
\end{align*}
as $N \rightarrow \infty$. To compute this quantity, we need to understand a bit better the properties of $U(\gamma)$. Recall that $U(\gamma)$ acts close to the black hole horizon as a translation along the null directions, 
\begin{equation}
   U(\gamma) = e^{-i\gamma P} = e^{-i\gamma P^{-}}, \text{ close to the horizon}
\end{equation}
   On the other hand, $\gamma = \frac{p_{+}}{M}e^{2\pi t_{N}}$ where $p_{+}$ is the momentum of the infalling perturbations, $B_{1,2}$. Since we can take $B_{1,2}$ to act late in the past in CFT$_{1}$/ far in the future in CFT$_{2}$, we take them to follow an almost null path very close to the horizon. Therefore,
   \begin{align*}
       U(\gamma) =& e^{-i\frac{P_{+}e^{2\pi t_{N}}}{M}P^{-}} \\
       =& 1-i\frac{e^{2\pi t_{N}}}{M}P_{+}P_{-} + \frac{1}{2!} \left(i\frac{e^{2\pi t_{N}}}{M}P_{+}P_{-}\right)^{2} + ...\\
       =& 1+ i \int\; dp_{+}dq_{-} \frac{e^{2\pi t_{N}}}{M} p_{+}q_{-}\ket{p_{+},q_{-}}\bra{p_{+},q_{-}} + \frac{1}{2!} \left(i \int\; dp_{+}dq_{-} \frac{e^{2\pi t_{N}}}{M} p_{+}q_{-}\ket{p_{+},q_{-}}\bra{p_{+},q_{-}}\right)^{2} +...
   \end{align*}
where $\ket{q_{-}}$ is the eigenstate of $-P^{-}$ since we are sending particles into the black hole so that they scatter with the infalling perturbations, $B_{1,2}$, while $P^{-}$ is translation operator into the future. An important point is that 
\begin{align*}
     \langle B^{'}_{1} U(-\gamma) D_{1}...\rangle &=  \langle B^{'}_{1}(1+ i \int\; dp_{+}dq_{-} \frac{e^{2\pi t_{N}}}{M} p_{+}q_{-}\ket{p_{+},q_{-}}\bra{p_{+},q_{-}} + ...)D_{1} ... \rangle \\
     &= \langle B^{'}_{1}D_{1} ... \rangle + i\langle B^{'}_{1} \int\; dp_{+}dq_{-} \frac{e^{2\pi t_{N}}}{M} p_{+}q_{-}\ket{p_{+},q_{-}}\bra{p_{+},q_{-}} D_{1} ... \rangle + ... \\
     &= \langle B^{'}_{1}D_{1} ... \rangle
\end{align*}
where $B_{1}^{'} = \sigma_{s}(B_{1})$.  The same can be said for the $U(-\gamma)$ acting directly on $B_{2}$. Thus,

\begin{align*}
    \langle B_{1}\;  \sigma_{s}( U^{\dagger}(\gamma)\;D_{1}\;U(2\gamma) \;D_{2}\; &U(-\gamma))\;B_{2})\rangle = \langle B^{'}_{1}\;  D_{1}\;U(2\gamma) \;D_{2}\; B^{'}_{2}\rangle \\
    &=  \int\; dp_{+}dq_{-} \text{exp}({2i \frac{e^{2\pi t_{N}}}{M}} p_{+}q_{-}) \langle B^{'}_{1}\;  D_{1}\ket{p_{+},q_{-}}\bra{p_{+},q_{-}}D_{2}\; B^{'}_{2}\rangle 
\end{align*}
 We can write it in a more familiar form,
 \begin{equation}
    \int\; dp_{+}dq_{-} \langle B^{'}_{1}\; \ket{p_{+}} \bra{p_{+}} B^{'}_{2}\rangle \left( e^{2i \frac{e^{2\pi t_{N}}}{M} p_{+}q_{-}} \langle D_{1}\ket{q_{-}}\bra{q_{-}}D_{2}\rangle \right) 
 \end{equation}
Bringing back in the smearing for $D_{1,2}$, we have,
\begin{equation}
    \Theta_{1}= ig  \int\; dp_{+}dq_{-} \langle B^{'}_{1}\; \ket{p_{+}} \bra{p_{+}} B^{'}_{2}\rangle \left( e^{2i \frac{e^{2\pi t_{N}}}{M} p_{+}q_{-}} \langle \tilde{D}_{1}\ket{q_{-}}\bra{q_{-}}\tilde{D}_{2}\rangle \right) + O(1/N)
\end{equation}
In addition,
\begin{align*}
    \Theta_{n} = & (ig)^{n} \langle B_{1}\;  \Delta^{-is} \left(U^{\dagger}(\gamma)\;\tilde{D}_{1}\;U(2\gamma) \;\tilde{D}_{2}\; U(-\gamma)\right)^{n} \Delta^{is}\;B_{2})\rangle + O(1/N)
\end{align*}
again leading to,
\begin{equation}
    \Theta_{n}= (ig)^{n}  \int\; dp_{+}dq_{-} \langle B^{'}_{1}\; \ket{p_{+}} \bra{p_{+}} B^{'}_{2}\rangle \left( e^{2i \frac{e^{2\pi t_{N}}}{M} p_{+}q_{-}} \langle \tilde{D}_{1}\ket{q_{-}}\bra{q_{-}}\tilde{D}_{2}\rangle \right)^{n} + O(1/N)
\end{equation}
Finally exponentiating the $\Theta_{n}$'s we get 
\begin{align*}
     C=&\langle e^{igL_{0}}\rangle \langle B_{1}(t_{1},\bar{x}) e^{igL_{N}}B_{2}(t_{2},\bar{x})\rangle + O(1/N) \\
       =& \langle e^{igL_{0}}\rangle  \int\; dp_{+}dq_{-} \langle B_{1}\; \ket{p_{+}} \bra{p_{+}} B_{2} \rangle \text{ exp}\left( e^{2i \frac{e^{2\pi t_{N}}}{M} p_{+}q_{-}} \langle \tilde{D}_{1}\ket{q_{-}}\bra{q_{-}}\tilde{D}_{2}\rangle \right) + O(1/N) \\
       =& \langle e^{igL_{0}}\rangle  \int\; dp_{+} \langle B_{1}\; \ket{p_{+}} \bra{p_{+}} B_{2} \rangle \text{ exp}\left(  \langle \tilde{D}_{1}e^{2i \frac{e^{2\pi t_{N}}}{M} p_{+}Q_{-}}\tilde{D}_{2}\rangle \right) + O(1/N)
\end{align*}
Thus reproducing \eqref{C} using a computation from a purely algebraic perspective. In particular we have derived that taking the large $N$ limit in the presence of a bounded function of $L_{N}$, taking the time ordering correctly, will correspond to a traversable wormhole in the bulk. The importance of the time ordering is implying that the bulk is traversable for perturbations in $\mathcal{M}$ that are far in the past; or for a shock wave that is far in the future, more precisely a scrambling time in the future from the perturbations.  Note that the $h$ prefactor in \eqref{C} is absorbed into the smearing of the $\tilde{D}_{1,2}$.

An important issue we want to emphasize is that this is not a computation done solely at the extreme of $N$ going to infinity limit, since in that case the bulk fluctuations are small to produce a strong back reaction on the eternal black hole background. Rather this computation should be taken as a different large $N$ limit where we have included certain operators that depend on $N$ in some way (more precisely the time at which they are applied on the thermofield double state depends on $N$) and the large $N$ limit of such modified state turns out to be the traversable eternal black hole. 

\section{Comments on the Hilbert space}\label{hilbert}

In this paper we have investigated the back-reaction of a negative energy shock wave on an eternal black hole geometry from an algebraic perspective. Even though naively one would need to use results from gravitational scattering computations to diagnose how probe perturbation react to the negative energy shock wave, we have found that the algebraic properties of the region close to the horizon, half sided modular inclusions and translations do reproduce the correct back-reaction on the geometry and path of the probes in this geometry. However, it would be important understand the details of the kind of Hilbert space one would have.

To this end we would like to quickly review how the extreme large $N$ Hilbert space of the eternal black hole is defined. The interesting kinds of operators of the CFT's in this limit are the single trace operators, as are here, and the subtracted single trace operators, that is single trace operators with there one point function subtracted, have a finite large N limit. If we denote the subtracted single trace operators by $\tilde{O}$, the large $N$ limit of the state $\ket{TFD}$ is defined as the state that gives the expectation value of any bounded function of the subtracted single trace operators as 

\begin{equation}
   \omega_{\Psi}(F(\tilde{O})) =  \bra{\Psi} F(\tilde{O})\ket{\Psi} = \lim_{N\rightarrow\infty}\bra{TFD}F(\tilde{O})\ket{TFD}
\end{equation}

The single trace operators could be either from the left or the right CFT's. If one tries to compute expectation values involving operators from both the left and right CFT's, there will be a non zero correlation even though the operators commute. 

Any other states in the Hilbert space are given by the action of the bounded functions of the single trace operators on $\ket{\Psi}$, $\ket{\Psi_{1}}= F(\tilde{O})\ket{\Psi}$ and $\ket{\Psi_{2}}= G(\tilde{O})\ket{\Psi}$ and the inner product between these two states is given by,

\begin{equation}
  \langle \Psi_{1}|\Psi_{2} \rangle =  \bra{\Psi} F^{\dagger}(\tilde{O})G(\tilde{O})\ket{\Psi} = \lim_{N\rightarrow\infty}\bra{TFD}F^{\dagger}(\tilde{O})G(\tilde{O})\ket{TFD}
\end{equation}

Taking the closure of the set of states like $\ket{\Psi_{1,2}}$ with respect to the above inner product gives the large $N$ Hilbert space.

Coming back our case of traversable wormhole, the interesting objects are again subtracted single trace operators both on the left and right CFT's. The essential point is the presence of the operator that couple the two CFT's $e^{igL_{N}}$ and produce a negative energy shock wave. The state that is created by the action of this operator at finite $N$ is, $$e^{igL_{N}}\ket{TFD}.$$ If one also acts with some bounded function of subtracted single trace operator from CFT-1, it would be $$e^{igL_{N}}F(O_{1})\ket{TFD}.$$ On the other hand if one acts with operator from CFT-2, one gets the state $$G(O_{2})e^{igL_{N}}\ket{TFD}.$$  

Thus, the extreme large $N$ state, $\ket{\Phi}$, for the traversable wormhole is the state that produces the following expectation values,
\begin{align}
    \omega_{\Phi}(F(O_{1})) &= \bra{\Phi}F(O_{1})\ket{\Phi} = \lim_{N\rightarrow\infty}\bra{TFD}e^{-igL_{N}}e^{igL_{N}}F(O_{1})\ket{TFD} = \lim_{N\rightarrow\infty}\bra{TFD}F(O_{1})\ket{TFD} \\
     \omega_{\Phi}(G(O_{2})) &= \bra{\Phi}G(O_{2})\ket{\Phi} = \lim_{N\rightarrow\infty}\bra{TFD}e^{-igL_{N}}G(O_{2})e^{igL_{N}}\ket{TFD}
\end{align}

Therefore, left-right correlation function is given by,
\begin{align}
     \omega_{\Phi}(G(O_{2})F(O_{1})) &= \bra{\Phi}G(O_{2})F(O_{1})\ket{\Phi} \\ 
     &= \lim_{N\rightarrow\infty}\bra{TFD}e^{-igL_{N}}G(O_{2})e^{igL_{N}}F(O_{1})\ket{TFD} 
\end{align}

where the last expression is the same as $C$, the correlator computed in the previous section, with $F(O_{1})$ being a CFT-1 operator and $G(O_{2})$ being the CFT-2 operator. As stated in the previous sections, this left-right correlation function encodes the signature for traversable wormhole geometry; the singularity that indicates causal connection between the left and right sides and the effect of a right unitary perturbation on the left CFT.  

The rest of the states are created by acting with bounded functions of subtracted single trace operators, $F(O)$ or $G(O)$, from either CFT-1, CFT-2 or both, on the state $\ket{\Phi}$. If $\ket{\Phi_{1}} = F(O)\ket{\Phi}$ and $\ket{\Phi_{2}} = G(O)\ket{\Phi}$, then their inner product is given by

\begin{equation}
    \langle\Phi_{1}|\Phi_{2}\rangle = \omega_{\Phi}(F^{\dagger}(O)G(O))
\end{equation}
Taking the closure of the space of states like $\ket{\Phi_{1,2}}$ with respect to the above inner product will give the large $N$ Hilbert space for traversable wormholes,

\begin{equation}
    \mathcal{H}_{\text{trav}} = \overline{\{F(O)\ket{\Phi}\}}.
\end{equation}

The full algebra of operators is constructed out of bounded functions of the subtracted single trace operators from the left and right CFT's, $F(O)$. We take the closure of this set by requiring that any Cauchy sequence of expectation values of these operators in the states in $\mathcal{H}_{\text{trav}}$, should have a limit point as the expectation value of some operator in the same set, leading to the full von Neumann algebra of operators for traversable wormhole.   

\section{Discussion and outlook}

An interesting future direction is the relationship of this algebraic discussion of traversable wormholes with quantum teleportation. It has indeed been understood early on \cite{Gao:2016bin,Maldacena:2017axo} that the Gao-Jafferis-Wall protocol can be interpreted as a quantum teleportation protocol from the boundary perspective. Therefore one also expects algebraic understanding quantum teleportation to play an important role. A recent work involving how information inside the black hole can become accessible from the radiation using index theory of type II$_{1}$ algebras and an operation called canonical endomorphism was discussed in \cite{vanderHeijden:2024tdk}. This procedure is a quantum teleportation protocol and it would be interesting to see how the use of conditional expectation and canonical endomorphism is connected to the work in this paper.

\section*{Acknowledgements}
I want to thank Amos Yarom, K. Papadodimas and the CERN-TH department for the support; and K. Papadodimas for useful discussions. This work is supported
by the Israel Science Foundation (ISF) Grants No. 3191/23 and No. 2916/24, and by the
United States-Israel Binational Science Foundation (BSF) Grant No. 2034142.
 
 \newpage
\appendix

\section{Algebra at infinity}\label{A}

In this appendix we discuss in detail the so called the algebra at infinity in the study of quasi local algebras. The central operator that produces negative energy shock wave is part of this algebra in the large $N$ limit since it is defined at $t_{N} \sim \frac{1}{2\pi} \text{log }N^{2}$. Quasi local algebras are quite important objects in quantum statistical mechanics and mathematical physics in general. We will mention some definitions and theorems (without proofs) that will clarify and motivate some of the discussions in the main body of the paper. We will be reviewing \cite{Bratteli:1979tw,Haag:1967sg,Lechner:2021fvy,Longo:1984zz} and more detailed discussion can be found in the same papers.

The usual formulation of quantum mechanics can not be applied to systems with infinite degrees of freedom in the infinite volume (thermodynamic) limit, without referring to quasi local algebras. These systems have unique properties that are not present in finite systems like phase transition. The standard discussion quantum many body systems involving the Fock's space of states, $\mathcal{H_{F}}$ constructed out of creation operators acting on a `no particle state', $\ket{\Omega}$, which is supposed to be annihilated by all annihilation operators. For a given square integrable function,$f$, $a(f)$ and $a^{\dagger}(f)$ act on $\mathcal{H_{F}}$ as annihilation and creation operators if
\begin{equation}
    a^{\dagger}(f) = \int d\bar{x} f(\bar{x}) a^{\dagger}(\bar{x}) ,\text{ and }\; a(f) = \int d\bar{x} f^{\star}(\bar{x}) a(\bar{x})
\end{equation}
where $a(\bar{x})$ and $a^{\dagger}(\bar{x})$ satisfy the canonical commutation(anti- commutation relations). Any state in $\mathcal{H_{F}}$ has a decreasing and decreasing probability for higher and higher number of particles, and in particular if one is interested in discussing infinite number of particles, $\mathcal{H_{F}}$ will prove useless. 

Restricting $f$ to be localized in some spatial subregion, $V$, we can construct the algebra of operators in the subregion $V$ using $a(\bar{x})$ and $a^{\dagger}(\bar{x})$ in that subregion, $\mathcal{Q}(V)$. The appropriate algebra for the thermodynamic limit is the norm closure of the union of all such algebras.

\begin{equation}
    \mathcal{Q}= \overline{{\cup_{V}(\mathcal{Q}(V))}}
\end{equation}

Note that since one is taking closure over the norm topology of the union, certain operators can not be included in the $\mathcal{Q}$, for instance the number operator or the Hamiltonian of the infinite system since they are truly infinite quantities in this limit. Therefore one has to restrict to what are called quasi local operators that are not actual global quantities. States over $\mathcal{Q}$ are taken as limits of the states on the finite volume algebras. Inherited from the canonical commutation or anti commutation relations is the property that $\mathcal{Q}(V_{1})$ commute(anticommute) with $\mathcal{Q}(V_{2})$ if $V_{1} \cap V_{2} $ is empty, and $V_{1}$ and $V_{2}$ are called disjoint.

One would like to generalize the above discussion as follows,

\begin{definition}
    
    A quasi local algebra is a $C^{\star}$ algebra $\mathcal{Q}$ and a net $\{\mathcal{Q}_{\alpha}\}_{\alpha \in I}$ of $C^{\star}$ algebras with an index set $I$ with a unique property called orthogonality relation and,

    \begin{enumerate}
        \item if $\alpha \geq \beta$ then $\mathcal{Q}_{\beta} \subseteq \mathcal{Q}_{\alpha}$;
        \item $\mathcal{Q} = \overline{ \cup_{\alpha} \mathcal{Q}_{\alpha}}$, where the bar is uniform closure;
        \item the algebras $\mathcal{Q}_{\alpha}$ have common identity;
        \item there is an automorphism $\gamma(\mathcal{Q}_{\alpha}) = \mathcal{Q}_{\alpha}$ such that $\gamma^{2} =1$ and 
        \begin{equation}
            [\mathcal{Q}_{\alpha}^{e},\mathcal{Q}_{\beta}^{e}] = \{0\}\; \text{ ,  } [\mathcal{Q}_{\alpha}^{e},\mathcal{Q}_{\beta}^{o}] = \{0\} \; \text{ , and } \{\mathcal{Q}_{\alpha}^{e},\mathcal{Q}_{\beta}^{o}\} = \{0\}
        \end{equation}
    \end{enumerate}
\end{definition}

whenever $\alpha$ and $\beta$ are orthogonal (which we will describe below), and where $\mathcal{Q}_{\alpha}^{e}, \mathcal{Q}_{\alpha}^{o} \subseteq \mathcal{Q}_{\alpha}$ are even and odd sub algebras of $\mathcal{Q}_{\alpha}$ with respect to $\gamma(.)$. For any element $A \in \mathcal{Q}_{\alpha}$ one has a unique decomposition, 
\begin{equation}
    A^{\pm} = \frac{A\pm \gamma(A)}{2}, \; \text{ with, } \gamma(A^{\pm}) = \pm A^{\pm}.
\end{equation}

$\mathcal{Q}_{\alpha}^{e}$ are construct out of the elements like $A^{+}$ corresponding to Bose statistics, while $\mathcal{Q}_{\alpha}^{o}$ are constructed out of $A^{-}$'s which correspond to Fermi statistics.

On the other hand, an index set $I$ is said to have an orthogonality relation (an intuitive generalization of disjointedness and union for the case of volumes discussed above), if there is a symmetric relation $\perp$ and $\vee$ (which corresponds to taking a union of two spatial sub regions) such that

\begin{enumerate}
    \item if $\alpha \in I$, then there is $\beta \in I$ with $\alpha \perp \beta$,
    \item if $\alpha \leq \beta$ and $\beta \perp \delta$, then $\alpha \perp \delta$,
    \item if $\alpha \perp \beta$ and $\alpha \perp \delta$ then there exists a $\nu$ such that $\alpha \perp \nu$ and $\nu \geq \beta, \delta$,
    \item if $\alpha,\beta \in I$, then $\alpha \vee \beta \in I$ and $\alpha,\beta \leq \alpha \vee \beta $
    \item in addition, if $\delta \geq \alpha,\beta$ then $\delta \geq \alpha \vee \beta$.
\end{enumerate}

Given a state $\omega$ over $\mathcal{Q}$, one can consider the GNS representation $(\mathcal{H}_{\omega}, \pi_{\omega}, \Omega_{\omega})$ that provide a useful description of the state $\omega$ and the von Neumann algebra of observables of this state, $\pi_{\omega}(\mathcal{Q})^{''}$. There are several important subalgebras of $\pi_{\omega}(\mathcal{Q})^{''}$ among which are the \emph{center, the commutant algebra} and \emph{the algebra at infinity}.  
\begin{definition}
    Given a GNS representation $(\mathcal{H}_{\omega}, \pi_{\omega}, \Omega_{\omega})$ with respect to some state $\omega$ on $\mathcal{Q}$,
    \begin{enumerate}
        \item The center $\mathcal{Q}_{ce,\omega}$ is a sub algebra of $\pi_{\omega}(\mathcal{Q})^{''}$ given by $$ \mathcal{Q}_{ce} = \pi_{\omega}(\mathcal{Q})^{''} \cap \pi_{\omega}(\mathcal{Q})^{'} $$ 
        \item The commutant algebra $\mathcal{Q}_{co,\omega}$ is a sub algebra of $\pi_{\omega}(\mathcal{Q})^{''}$ given by $$ \mathcal{Q}_{co} = \cap_{\alpha\in I}\;(\pi_{\omega}(\mathcal{Q}_{\alpha})^{'} \cap \pi_{\omega}(\mathcal{Q}))^{''} $$ 
        \item The algebra at infinity $\mathcal{Q}_{\infty,\omega}$ is a sub algebra of $\pi_{\omega}(\mathcal{Q})^{''}$ given by $$ \mathcal{Q}_{\infty} = \cap_{\alpha\in I}\;(\cup_{\beta \perp \alpha}\;\pi_{\omega}(\mathcal{Q}_{\beta}))^{''} $$  
    \end{enumerate}
\end{definition}

The algebra $\mathcal{Q}_{\infty,\omega}$ is, for the intuitive example discussed at the beginning of the appendix, is really the algebra of operators that are localized at the spatial infinity of the system and therefore can be measured at infinity. It also have interesting relationship with cluster decomposition principle in `reasonable' quantum field theories. Since the operators that are satisfying the Fermi statistics present certain complications, here we just focus on the case where $\gamma = 1$.

\begin{theorem}
    For a given state $\omega$ on $\mathcal{Q}$, the algebra at infinity $\mathcal{Q}_{\infty,\omega}$ is a sub algebra of $\mathcal{Q}_{ce,\omega}$ and the following statement are equivalent;
    \begin{enumerate}
        \item $\mathcal{Q}_{\infty,\omega}$ = $\{\mathbf{C}\mathrm{I}\}$, i.e, multiples of identity,
        \item For any $A \in \mathcal{Q}$, there is $\alpha \in I$ such that $$|\omega(AB) - \omega(A)\omega(B)| \leq || \pi_{\omega}(B)||,$$ for all $B\in \mathcal{Q}_{\beta}$ for any $\beta \perp \alpha$,
        \item For any $A \in \mathcal{Q}$, there is $\alpha \in I$ such that $$|\omega(AB) - \omega(A)\omega(B)| \leq (\omega(BB^{\star})+\omega(B^{\star}B))^{1/2},$$ for all $B\in \mathcal{Q}_{\beta}$ for any $\beta \perp \alpha$,
        
    \end{enumerate}
\end{theorem}

There are several places in the literature where algebra at infinity arises some of which we will discuss here. 

\begin{definition}
    Let $(\mathcal{M,N},\Omega)$, where $\mathcal{M,N}$ are von Neumann algebras, be a standard inclusion that is $\mathcal{N}\subset \mathcal{M}$ with $\Omega $ is cyclic and separating for $\mathcal{M,N, M \cap N^{'}}$. Let $J_{\mathcal{M}},J_{\mathcal{N}}$ be modular conjugations for $\mathcal{M,N}$ with respect to $\Omega$, then $$\Gamma (a) = J_{\mathcal{N}}J_{\mathcal{M}} a J_{\mathcal{M}}J_{\mathcal{N}}, \; \text{ for } a \in \mathcal{M} $$ defines a unitary map from $\mathcal{M}$ to $\mathcal{N}$.

    Consider a multiple application of $\Gamma(.)$ to define a chain $\mathcal{A}_{k}$ given by; $\mathcal{A}_{k} = \Gamma^{k}(\mathcal{M})$ when $k$ is even and $\mathcal{A}_{k} = \Gamma^{k}(\mathcal{N})$ when $k$ is odd.

    Then we have 
    \begin{equation}
        \mathcal{A}_{0}= \mathcal{M} \supset \mathcal{A}_{1}=\mathcal{N} \supset \mathcal{A}_{2} \supset ...
    \end{equation}

    Finally the algebra at infinity defined to be $$\mathcal{A} = \cap_{k}\mathcal{A}_{k}$$
\end{definition}

$\mathcal{A}$ has some interesting properties for instance,
\begin{lemma}
If $\mathcal{A}=\cap_{k}\mathcal{A}_{k}$ is the algebra at infinity, then  
\begin{enumerate}
    \item If $\mathcal{M}^{\Gamma} = \{a \in \mathcal{M} : \Gamma(a)=a\}$, the $\mathcal{M}^{\Gamma} = \mathcal{A}$.
    \item Let $Z(\mathcal{M})$ and $Z(\mathcal{N})$ denote the centers of $\mathcal{M}$ and $\mathcal{N}$. If $Z(\mathcal{M}) \wedge Z(\mathcal{N}) = \{\mathbf{C}\mathrm{I}\}$, then $\mathcal{A}$ is a type III$_{1}$ algebra or $\mathcal{A} =\mathbf{C}\mathrm{I} $.
\end{enumerate}
\end{lemma}

\subsection{Algebra at infinity and Half sided modular inclusions}

Another instructive example is the case where we consider half sided modular inclusions, i.e, $(\mathcal{M,N},\Omega)$, where $\mathcal{M,N}$ are von Neumann algebras such that $\mathcal{N}\subset \mathcal{M}$ with $\Omega $ is cyclic and separating for $\mathcal{M,N}$, and $$\sigma_{t}(\mathcal{N})=\Delta^{-it}\mathcal{N}\Delta^{it} \subset \mathcal{N}$$ for $t\geq 0$ (or $t\leq 0$), where $\Delta$ is the modular operator for $\mathcal{M}$. If $J$ is the modular conjugation operator for $\mathcal{M}$, then we will also have $$\sigma_{t}(J\mathcal{N}J)=\Delta^{-it}J\mathcal{N}J\Delta^{it} \subset J\mathcal{N}J$$ for $t\leq 0$ (or $t\geq 0$).

In the AdS/CFT context, $\mathcal{M}$ can be thought of as the large $N$ limit von Neumann algebra of operators corresponding to the right CFT. The algebra $\mathcal{N}$ is a sub algebra of $\mathcal{M}$, while $J\mathcal{N}J$ is a sub algebra of $\mathcal{M}^{'}$. Without loss of generality, one chooses $\mathcal{N}$ to be such that the half sided modular inclusion is for $t\geq0$. One then defines the left(right) algebra at infinity by

\begin{equation}
    \mathcal{A}_{\infty,right} = \cap_{t>0}\; \sigma_{t}(\mathcal{N})\; \text{ and, }  \mathcal{A}_{\infty,left} = \cap_{t<0}\; \sigma_{t}(J\mathcal{N}J)
\end{equation}

For the cases where $\mathcal{M,N}$ are factors and the vacuum $\Omega$ is unique $\mathcal{A}_{\infty,right}=\mathcal{A}_{\infty,left} = \{\mathbf{C}\mathrm{I}\}$. However, for the cases where there is a non-trivial center for $\mathcal{M,N}$, for instance in the example of the extreme large $N$ limit of the canonical ensemble \cite{Leutheusser:2021frk, Leutheusser:2022bgi, Witten:2021unn}, the left and right algebras at infinity solely consists of the algebra of conserved charges that commute with the modular (time in the boundary theory) translations. The same can be said for the extreme large $N$ limit of the microcanonical black hole \cite{Chandrasekaran:2022eqq}. 

On the other hand, in the same context a different algebra at infinity was introduced in the \cite{Lechner:2021fvy}, which was the motivation to consider the yet different algebra at infinity considered in section \ref{alg tra worm}.

\begin{definition}
    Let $(\mathcal{N \subset M},\Omega)$ be a half sided modular inclusion with associated half sided modular translations, a strongly continuous unitary group, $U$, with positive generator, $P$, and;
    \begin{enumerate}
        \item $U(x)\ket{\Omega} = \ket{\Omega}$
        \item $\alpha_{x}(\mathcal{M}) := U^{-1}(x) \mathcal{M}U(x) \subseteq \mathcal{M}$, for $x\geq0$.
    \end{enumerate}
Then, for a bounded interval $(a,b) \in \mathbf{R}$, we define $$\mathcal{A}(a,b) = \alpha_{a}(\mathcal{M}) \cap \alpha_{b}(\mathcal{M})^{'}.$$

In particular, for any bounded interval $\mathcal{I}$ in $\mathbf{R}$, one has an associated algebra $\mathcal{A(I)}$. 

In addition, we define the algebra at infinity,$$\mathcal{A}_{\infty} = \cap_{\mathcal{I}} \mathcal{A(I)}^{'}$$

Alternatively,
$$\mathcal{A}_{\infty} = \cap_{x>0}(\alpha_{x}(\mathcal{M}) \cap \alpha_{-x}(\mathcal{M})^{'}) = \cap_{t>0}\; \sigma_{t}(\mathcal{N} \vee J\mathcal{N}J) $$

\end{definition}

As emphasized in \cite{Lechner:2021fvy}, even though $\mathcal{A}_{\infty,right}$ and $\mathcal{A}_{\infty,left}$ are trivial under the assumptions, $\mathcal{A}_{\infty}$ can be highly non-trivial.

\begin{theorem}
    Let $\mathcal{M}$ and $\mathcal{N}$ be von Neumann algebras acting on the Hilbert space $\mathcal{H}$ and $A \in \mathcal{N}\vee J\mathcal{N}J$ be such that $w-\lim_{t\rightarrow\infty} \sigma_{t}(A)= L $. Then,
    \begin{enumerate}
        \item $L \in \mathcal{A}_{\infty}$,
        \item $[L,\Delta^{it}]=0$ for all $t \in \mathbf{R}$,
        \item $L\ket{\Omega} = \omega(A)\ket{\Omega}$
    \end{enumerate}
\end{theorem}

\newpage

\bibliographystyle{JHEP}
\bibliography{biblio}

\end{document}